\RequirePackage{luatex85}
\documentclass[11pt]{article}
\usepackage{jheppub}
\usepackage{amsmath}
\usepackage{amssymb}

\usepackage{tikz}
\usetikzlibrary{intersections, calc,shapes.geometric,positioning,patterns,arrows,decorations.pathreplacing, matrix,decorations.pathmorphing}
\usepackage{tkz-euclide} 
\usetkzobj{all} 
\usepackage{tikz-3dplot}
\usepackage[all]{xy}

\usepackage{color}

\def\CB{{\cal B}}
\def\CC{{\cal C}}
\def\CD{{\cal D}}

\def\CM{{\cal M}}
\def\CN{{\cal N}}
\def\CO{{\cal O}}

\def\BH{\mathbb{H}}
\def\BR{\mathbb{R}}

\def\BZ{\mathbb{Z}}

\newcommand{\Mdm}{\mathcal{M}^{(d,m)}}

\title{Operator product expansion for conformal defects}

\author[a]{Masayuki Fukuda,}
\author[a,b]{Nozomu Kobayashi}
\author[a]{and Tatsuma Nishioka}

\affiliation[a]{Department of Physics, Faculty of Science,
The University of Tokyo,\\
Bunkyo-ku, Tokyo 113-0033, Japan}
\affiliation[b]{Kavli Institute for the Physics and Mathematics of the Universe (WPI), \\
The University of Tokyo Institutes for Advanced Study, The University of Tokyo, \\
Kashiwa, Chiba 277-8583, Japan}

\abstract{
We study the operator product expansion (OPE) for scalar conformal defects of any codimension in CFT.
The OPE for defects is decomposed into ``defect OPE blocks", the irreducible representations of the conformal group, each of which packages the contribution from a primary operator and its descendants.
We use the shadow formalism to deduce an integral representation of the defect OPE blocks.
They are shown to obey a set of constraint equations that can be regarded as equations of motion for a scalar field propagating on the moduli space of the defects.
By employing the Radon transform between the AdS space and the moduli space, we obtain a formula of constructing an AdS scalar field from the defect OPE block for a conformal defect of any codimension in a scalar representation of the conformal group, which turns out to be the Euclidean version of the HKLL formula.
We also introduce a duality between conformal defects of different codimensions and prove the equivalence between the defect OPE block for codimension-two defects and the OPE block for a pair of local operators.
}

\preprint{UT-17-33, IPMU17-150}

\begin{document}
\maketitle

\section{Introduction}
Non-local objects such as loop and surface operators are ubiquitous in quantum field theories, and provide invaluable knowledge about the status of the theories that are inaccessible by other means.
\emph{Defects} collectively mean non-local operators of various (co)dimensions.
They are less understood than local operators except when they are given explicit representations by fundamental local fields, especially because they are often characterized as boundary conditions on their supports as most exemplified by 't Hooft loops.\footnote{There are several ways defining and constructing defects.
See e.g.\,\cite{Gukov:2014gja} for a review.
}
Among them a particularly important class is conformal defects preserving a part of the conformal symmetry in CFT and allowing better control of determining correlation functions by the residual symmetry.
For instance, boundary and interface CFTs are examples of codimension-one conformal defects \cite{Cardy:1984bb,McAvity:1995zd}.
Twist operators associated to entanglement entropy are an intriguing example of  codimension-two defects living on an entangling surface in CFT \cite{Calabrese:2004eu,Cardy:2013nua,Hung:2014npa,Bianchi:2015liz}.

In recent years there has been an increasing interest in defect CFTs mainly due to the development of the conformal bootstrap program that aims to constrain the dynamics of CFT by a set of data defining defects \cite{Liendo:2012hy,Gaiotto:2013nva,Gliozzi:2015qsa}.
Systematic studies for conformal defects of arbitrary codimension have been undertaken in \cite{Billo:2016cpy,Gadde:2016fbj} where the embedding space formalism \cite{Dirac:1936fq,Mack:1969rr,Weinberg:2010fx,Costa:2011mg} was extended to deal with the correlation functions of local operators in the presence of defects.
In this formalism, the position of a conformal defect of codimension-$m$ in CFT$_d$ is given by the intersection of a codimension-$m$ hypersurface and the null projective cone in the embedding space $\BR^{d+1,1}$, preserving the $SO(m)\times SO(d+1-m,1)$ subgroup of the entire conformal symmetry $SO(d+1,1)$.
While there are bulk local operators $\CO$ in the bulk CFT, 
the conformal defect can also accommodate defect local operators $\hat\CO$ that transform under the conformal group $SO(d+1-m,1)$ on the codimension-$m$ hypersurface with ``flavor" symmetry $SO(m)$.
Defect primary operators $\hat \CO_n$, which span a basis of the defect local operators, are classified by the irreducible representations (which we denote $n$ in the subscript of $\hat\CO$) of the subgroup $SO(m)\times SO(d-m) \times SO(1,1)$, i.e., two irreducible representations of $SO(m)$ and $SO(d-m)$, and the conformal dimension $\hat\Delta$, as the bulk primary fields $\CO_n$ are labeled by the irreducible representation of $SO(d)$ and the conformal dimension $\Delta$.

The bulk and defect local operators are not independent of each other.
To inspect their relation, it is tempting to expand a bulk local operator $\CO(x^a, x^i)$ by the defect primary operators $\hat \CO_n(x^a)$ and their descendants,
\begin{align}\label{Bulk-to-Defect_OPE}
    		\CO(x^a, x^i) = \sum_{n} b_{\CO\,\hat \CO_n}\, |x^i|^{\hat\Delta_n - \Delta}\, \hat\CO_n(x^a) + \text{(descendants)}\ ,
\end{align}
where $x^a$ and $x^i$ are the coordinates parallel and orthogonal to the defect, respectively.
This type of Operator Product Expansion (OPE) is called the \emph{bulk-to-defect OPE}.
The OPE coefficients $b_{\CO\,\hat \CO_n}$ fix the relation between the bulk and the defect local operators, and further provide the information about and the constraints on the OPE coefficients of the bulk operators \cite{McAvity:1995zd,Liendo:2012hy,Billo:2016cpy}.
In applying the bulk-to-defect OPE, the bulk operator $\CO$ is supposed to be close to the defect.
Hence we can find a (quantization) sphere surrounding $\CO$ and intersecting the defect, and associate to the sphere a state specified by the bulk operator $\CO$ and the defect.
The bulk-to-defect OPE \eqref{Bulk-to-Defect_OPE} states that the state is described in the radial quantization by a set of the defect local operators when the sphere is shrunk to a point on the defect by the scale transformation.
On the other hand, when the bulk field is far from the defect, it is more convenient, if it is spherical, to adopt a quantization surface enclosing both $\CO$ and the defect.
In this case, the state is rather expanded by a set of the bulk operators on a point where the quantization surface shrinks by the scale transformation.
For example, consider a spherical defect of radius $R$ located on the origin, and choose the identity operator as the bulk operator.
Then the aforementioned argument amounts to the expansion of the defect $\CD$ itself by the bulk operators,
\begin{align}\label{defect_OPE_Intro}
    		\CD = \sum_n c_{\CO_n}\, R^{\Delta_n}\, \CO_n(0) + \text{(descendants)}\ .
\end{align}
The OPE for defects of this type was applied for studying loop and surface operators in \cite{Berenstein:1998ij,Corrado:1999pi,Chen:2007zzr,Gomis:2009xg} and entanglement entropy in \cite{Cardy:2013nua}, and has been revisited recently from a more abstract viewpoint in \cite{Gadde:2016fbj}.
In this paper we mean by the \emph{defect OPE} the expansion \eqref{defect_OPE_Intro} of conformal defects.\footnote{The expansion \eqref{defect_OPE_Intro} is named the defect expansion in \cite{Gadde:2016fbj} to distinguish from the bulk-to-defect OPE that is sometimes called the defect OPE in literature.
}
Now we have two complimentary ways, using either the bulk-to-defect OPE \eqref{Bulk-to-Defect_OPE} or the defect OPE \eqref{defect_OPE_Intro}, to evaluate the one-point function $\langle \CO(x)\rangle_\CD$ of a bulk operator in the presence of a conformal defect.
Comparison between the two results shows that the defect OPE coefficients $c_\CO$ are proportional to $b_{\CO \hat {\bf 1}}$ \cite{Gadde:2016fbj}.\footnote{
More generally, a defect local operator $\hat\CO$ would allow the expansion,
    \begin{align}
    		\hat\CO (y^a) = \sum_j c_{\CO_j\hat \CO}\, f(y^a, z^a, z^i, \partial_{z^a}, \partial_{z^i})\, \CO_j(z^a, z^i) \ .
    	\end{align}
    The OPE coefficients $b_{\CO\, \hat\CO}$ and $c_{\CO\,\hat \CO}$ in the two OPEs are related through the three-point function $\langle \CO(x^a, x^i)\, \hat \CO(y^a)\rangle$, while the position $(z^a, z^i)$ and the unknown function $f(y^a, z^a, z^i, \partial_{z^a}, \partial_{z^i})$ should be fixed so that they give the same result.
}
Hence the defect OPE contains the same amount of information about the defect as the bulk-to-defect OPE does while the former has been less explored than the latter so far.

The objective of the present paper is twofold.
First, we want to investigate the universal structure of the defect OPE  for conformal defects that can be determined by the conformal symmetry.
To this end, it is neat to collect the contribution from a conformal multiplet of a primary operator $\CO_n$ in the defect OPE \eqref{defect_OPE_Intro} into a non-local function $\CB[\CO_n]$, which we call the \emph{defect OPE block} of $\CO_n$, that transforms in the same way as $\CO_n$ under the conformal group,
\begin{align}
    		\CD = \sum_n\, \CB[\CO_n]\ .
\end{align}
In principle, the descendant terms in the block can be fixed term by term so that correlation functions of the bulk operators with the defect are invariant under the residual conformal symmetry.
This approach has been often adopted in literature though, 
it obscures some fundamental structures of the defect OPE blocks. 
We would rather employ the shadow formalism developed by \cite{Ferrara:1972ay,Ferrara:1973vz,Ferrara:1972xe,Ferrara:1972uq,Mack:1975jr,Mack:1976pa} and implemented in the embedding space by \cite{SimmonsDuffin:2012uy}, which enables us to decompose the defect OPE into the blocks $\CB[\CO_n]$ by the projector $|\CO_n|$ onto the conformal multiplet of $\CO_n$.
The projector is schematically given by the dimensionless integral of the form,
\begin{align}
	 \int d^d x \, |\CO_n(x) \rangle\, \langle \tilde\CO_n (x)| \ ,
\end{align}
where $\tilde\CO_n (x)$ is the shadow operator defined by a non-local function of $\CO_n(x)$ with dimension $d-\Delta$ when $\CO_n(x)$ has dimension $\Delta$ in CFT$_d$.
Armed with the projectors and the spectral decomposition of the identity operator ${\bf 1} = \sum_n\, |\CO_n|$, we derive the defect OPE block in the integral representation,
\begin{align}\label{DOPE_Block}
	\CB[\CO_n] = \int d^d x \, \langle \CO_n(x) \rangle_\CD\, \tilde \CO_n(x)\ ,
\end{align}
where the one-point function $\langle \CO_n(x) \rangle_\CD$ does not necessarily vanish in contrast to the case of CFT without defects.
We will focus on the blocks for spinning operators and examine a set of constraints imposed on them by the conformal symmetry.

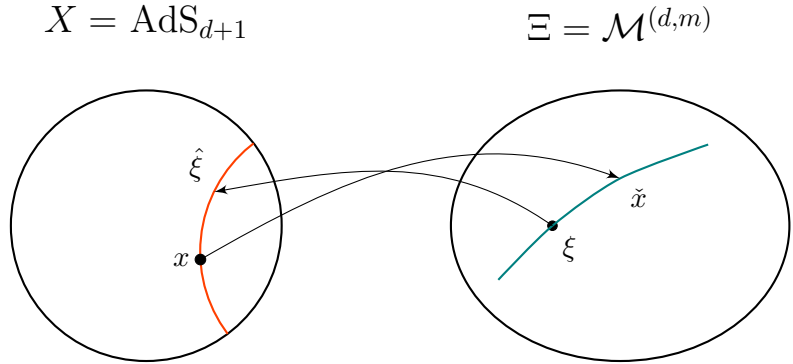
\begin{figure}[t]
	\centering
    \begin{tikzpicture}[>=latex',scale=0.9]
    	
        \draw[thick] (4,0) circle (2.5cm and 2cm);
        \coordinate (xi) at (3,0);
        \coordinate (xc) at (4,0.7);
		\filldraw[black] (xi) circle (2pt) node[below right] {$\xi$}; 
        \draw[thick, blue!50!green] plot [smooth] coordinates {(2.2, -0.8) (xi) (xc) (5.3,1.2)};
        \node[below right] at (xc) {$\check{x}$};
        
		\tkzDefPoint(-3,0){O}
  		\tkzDefPoint(-1,0){A}
  		\tkzDrawCircle[thick](O,A)
  		\tkzDefPoint(-2,0.5){xih}
  		\tkzDefPoint(-2.2,-0.5){x}
        \begin{scope}
  		\tkzClipCircle(O,A)
  		\tkzDrawCircle[orthogonal through=xih and x, thick, color=red!50!orange](O,A)
        \end{scope}
  		\tkzDrawPoints[color=black,fill=black,size=5](x)
  		\tkzLabelPoints[left](x)
        \node[above left] at (xih) {$\hat \xi$};
        
        \draw[->] (xi) to [out=145, in=10] (xih);
        \draw[->] (x) to [out = 30, in=160] (xc);
        
        \node at (-3,3) {\Large $X =$ AdS$_{d+1}$};
        \node at (4,3) {\Large $\Xi = \CM^{(d,m)}$};
    \end{tikzpicture}
    \caption{The Radon transform between the AdS space and the moduli space $\CM^{(d,m)}$ of conformal defects}
    \label{fig:Radon_AdSCFT}
\end{figure}

The second objective is leveraging the implications of the defect OPE blocks for the AdS/CFT correspondence.
Our starting point is to view $\CB[\CO_n]$ as a scalar field propagating on the moduli space $\CM^{(d,m)}$ of codimension-$m$ conformal defects.
Having in mind the AdS/CFT correspondence, we identify $\CM^{(d,m)}$ with the moduli space of totally geodesic codimension-$m$ hypersurfaces in the AdS$_{d+1}$ space, both having the same coset space structure, $SO(d+1,1)/SO(m)\times SO(d+1-m,1)$.
Under this identification we are able to map a scalar function $\phi(x)$ on the AdS$_{d+1}$ space to a scalar function $\hat \phi (\xi)$ on the moduli space by the Radon transform that smears $\phi(x)$ on a hypersurface $\hat\xi$ corresponding to a point $\xi \in \CM^{(d,m)}$ (see figure \ref{fig:Radon_AdSCFT}).
The original function $\phi(x)$ can be reconstructed from $\hat\phi(\xi)$ by the inversion formula of the Radon transform,\footnote{The inversion formula was applied in \cite{Lin:2014hva} to extract the energy density in an asymptotically AdS space from the relative entropy between the ground state in CFT and an excited state dual to the bulk space.} which roughly speaking integrates $\hat\phi(\xi)$ over the set of all hypersurfaces $\hat\xi$ at some distance $p$ from the point $x$ and then smears it over $p$ with some weight.
Thus the inversion formula provides us a concrete procedure to build an AdS scalar field from the set of the defect OPE blocks in CFT by designating $\hat\phi = \CB[\CO_n]$.
Moreover, the intertwining property of the Radon transform plays an important role in translating the constraints on the blocks into the equation of motion in the AdS space.
Combined with the integral representation of the block \eqref{DOPE_Block}, we argue the resulting expression of the AdS scalar field agrees with the Euclidean version of the HKLL formula \cite{Hamilton:2006fh}.

Some of the methods and ideas used in this paper are not completely new,
and owe to the earlier works \cite{Czech:2016xec,deBoer:2016pqk} that present a dictionary connecting the AdS scalar field with the entanglement entropy across a sphere in CFT through the Ryu-Takayanagi formula \cite{Ryu:2006bv,Ryu:2006ef}.\footnote{The dictionary was extended to boundary CFT by \cite{Karch:2017fuh} where the structure of the bulk-to-defect OPE \eqref{Bulk-to-Defect_OPE} was investigated. See also \cite{Herzog:2017xha,Karch:2017wgy}.
}
In particular they associate the OPE of local operators at two timelike separated points to an integration over the causal diamond of a codimension-two surface located at the intersection of the light cones passing through the two points.
The integral, termed by the OPE block, is a bilocal function containing contributions from a conformal multiplet to the OPE.
We show the OPE block is a special case of the defect OPE block for a codimension-two defect, and reproduce the known properties from scratch.
We also clarify the interplay between the codimension-two defect and the pair of points on the tips of the causal diamond by introducing a duality between two conformal defects of codimension-$m$ and codimension-$(d+2-m)$.

For the sake of simplicity, we will consider only scalar defects throughout this paper, though our derivation for the defect OPE block should apply equally to spinning defects.
We will also ignore conformal anomalies in the presence of defects, which have attracted rising attention in the recent studies of boundary and defect anomalies \cite{Herzog:2015ioa,Fursaev:2015wpa,Solodukhin:2015eca,Jensen:2015swa,Herzog:2017xha,Herzog:2017kkj}.
Some of them are related to Graham-Witten anomaly that can appear in defect CFT when the dimension of defects is even \cite{Graham:1999pm}, and are important to understand the universal part of entanglement entropy in CFT \cite{Solodukhin:2008dh,Fursaev:2013fta,Fursaev:2016inw}.
These will be left as interesting future problems.

This paper is organized as follows.
Section \ref{ss:ConformalDefect} contains a review of conformal defects in the embedding space formalism and a discussion on the duality between defects of different codimensions.
The codimension-two case is treated separately for later use to make contact with the OPE block of local operators.
In section \ref{ss:Defect_OPE_block}, we define the defect OPE blocks and determine their forms by exploiting the shadow formalism.
A monodromy condition is imposed on the block to remove the contribution of the shadow operators as in the case of the conformal block.
We examine the structure of and constraints on the defect OPE blocks in detail, which are compared with the OPE blocks when codimension is two.
Section \ref{ss:Radon} elucidates the relation of the physics between the AdS space and conformal defects in CFT.
The Radon transform between the AdS space and the moduli space of conformal defects fits for this purpose.
Using the inversion formula of the Radon transform, we derive a formula constructing an AdS scalar field from the defect OPE blocks in the scalar representation.
The equivalence of our formula to the HKLL formula is argued by adopting the integral representation of the block.
In section \ref{ss:Discussion}, we discuss possible applications of our results and several open issues left for further investigation.

\section{Conformal defects}\label{ss:ConformalDefect}
The conformal symmetry $SO(d+1,1)$, acting non-linearly on fields in $d$ dimensions, can be realized linearly in the embedding space $\BR^{d+1,1}$ \cite{Dirac:1936fq,Mack:1969rr,Ferrara:1973eg}, which makes it easy to determine the correlators of higher spin fields \cite{Weinberg:2010fx,Costa:2011mg}.
This formalism has been employed and expanded to describing conformal defects of any codimension and their correlators with local operators in \cite{Billo:2016cpy,Gadde:2016fbj}.
In this section, we summarize basic tools necessary for proceeding to the main part of this paper in the later sections.
Section \ref{ss:duality} introduces a new duality between conformal defects of different codimensions, which will play a key role in relating our defect OPE block and the OPE block in section \ref{ss:OPE_block}.

\subsection{Embedding space formalism}\label{ss:Embedding}
We consider CFT in a Euclidean space $\BR^d$ for simplicity, but the discussion below can be extended straightforwardly to a Lorentzian CFT as well.
The Euclidean conformal group is $SO(d+1,1)$ and the generators of this group act on a primary operator $\CO(x)$ in a flat $d$-dimensional space $\BR^d$ as follows:
\begin{align}\label{Conformal_Algebra}
\begin{aligned}
\hat M_{\mu \nu}\, \CO(x) &= (x_\mu \partial_{\nu} - x_{\nu} \partial_{\mu}+ S_{\mu \nu})\,\CO(x) \ ,\\
\hat P_{\mu}\,\CO(x) &= \partial_{\mu} \CO(x) \ , \\
\hat K_{\mu}\,\CO(x) &= (2\Delta\, x_{\mu}+2x^{\nu} S_{\nu \mu} -x^2\partial_{\mu} + 2x_{\mu}x^{\nu}\partial_{\nu})\,\CO(x) \ ,\\
\hat D\,\CO(x) & = (x^{\mu}\partial_{\mu}+\Delta)\,\CO(x)\ ,
\end{aligned}
\end{align}
where $\hat M_{\mu \nu}$ are the generators of the rotation group $SO(d)$, $\hat P_{\mu}$ the translation, $\hat K_{\mu}$ the special conformal transformation, and $D$ the dilatation.
$\Delta$ and $S_{\mu \nu}$ are the conformal dimension and the spin matrix associated with $\CO(x)$, respectively.

While these actions of the conformal group are non-linear we can realize them linearly by enlarging a space to the embedding space $\BR^{d+1,1}$
and restricting a point $X$ in the embedding space to the projective null cone, $ X^2 = 0 $, with an identification $X \sim \lambda X$, $\lambda \in \BR$. 
The last condition means there is a gauge redundancy of the rescaling in the embedding space.

Using the light cone coordinates $X^A = (X^+,X^-,X^i),\, (i =1,\cdots,d)$, 
the $SO(d+1,1)$ invariant inner product in the embedding space is defined by 
\begin{align}
	X \cdot Y = -\frac{1}{2}(X^+ Y^- + X^- Y^+) + \sum_{i=1}^d X^i Y^i \ .
\end{align}
The null cone condition $X^2 = 0$ is satisfied by choosing the point as $X^A =(\alpha,x^2 / \alpha,x^i)$ where $\alpha$ is an arbitrary constant.
We can use the gauge redundancy $X \sim \lambda X$ to set $\alpha = 1$, which is called the Poincar{\'e} section.
Then the coordinates of the physical space $\BR^d$ are given by $x^i$ $(i=1,\cdots, d)$ in the Poincar{\'e} section, $X=(1,x^2,x^i)$.

The conformal algebra \eqref{Conformal_Algebra} can be packed in the embedding space into the generators $\hat J_{AB}$ $(A,B = 1,\cdots, d+2)$ obeying the commutation relations,
\begin{align}
	[\hat J_{AB} \ , \hat J_{CD} ] = \eta_{AD}\, \hat J_{BC} + \eta_{BC}\,\hat J_{AD} - \eta_{AD}\,\hat J_{BC} - \eta_{BD}\,\hat J_{AC} \ ,
\end{align}
with the metric $\eta_{AB} \equiv \text{diag} (1, 1, \cdots, 1, -1)$, which is used when the indices of a vector is raised and lowered such that $X_A\equiv \eta_{AB}X^B$.
It will be convenient to decompose the generator into the orbital part $\hat L_{AB}$ and the spin part $\hat S_{AB}$ commuting with each other,
\begin{align}
	\hat J_{AB} = \hat L_{AB} + \hat S_{AB} \ .
\end{align}
When acted on a local operator at a position $X$ in the embedding space  the orbital part $\hat L_{AB}$ is realized as a derivative operator,\footnote{We use the hatted notation for the conformal generators $\hat L_{AB}$ to distinguish from their representations as differential operators $L_{AB}$.
In our convention the generators are anti-hermitian, $\hat L_{AB}^\dagger = - \hat L_{AB}$.
}
\begin{align}\label{L_X}
	L_{AB}(X) = X_{A} \frac{\partial}{\partial X^B} - X_{B} \frac{\partial}{\partial X^A} \ .
\end{align}
The spin part $\hat S_{AB}$ is the uplift of the spin matrix $S_{\mu\nu}$ into the embedding space \cite{Weinberg:2010fx}.

\subsection{Correlators in embedding space}
The embedding space formalism manifests the conformal invariance of correlation functions, and is particularly useful to investigate correlation functions of spinning operators \cite{Costa:2011mg}.
The extended approach for treating operators in general tensor representations was developed in \cite{Costa:2014rya,Costa:2016hju}, but we will not use it in this paper.

First, let us consider the two-point function of a scalar field $\phi(x)$ with dimension $\Delta$.
We uplift $\phi(x)$ to a scalar field $\Phi(X)$ in the embedding space.
The two-point function $\langle \Phi(X_1) \Phi(X_2) \rangle$ should be a scalar, but we can construct only one scalar invariant $X_1\cdot X_2$ from the vectors $X_1$ and $X_2$.
The two-point function should be a function of both $X_1$ and $X_2$ with degree $-\Delta$, and it ends up with the following unique form,
\begin{align}
 \langle \Phi(X_1) \Phi(X_2) \rangle = \frac{1}{(X_{12})^{\Delta}} \ ,
\end{align}
where we introduced the shorthand notation $X_{ij} \equiv -2 X_i\cdot X_j$ that reduces to the distance squared between the two points $(x_i - x_j)^2$ in the Poincar{\'e} section.
The normalization constant is chosen so as to reproduce the conventional form $1/(x_1 - x_2)^{2\Delta}$.

In order to deal with a traceless symmetric tensor $f_{a_1\cdots a_l}(x)$ of dimension $\Delta$, we consider a tensor field $F_{A_1 \cdots  A_l}(X)$ in the embedding space satisfying the following conditions:
\begin{itemize}
\item Homogeneous of degree $-\Delta$: $F_{A_1 \cdots  A_l}(\lambda X) = \lambda^{-\Delta}F_{A_1 \cdots  A_l}(X)$\ ,
\item Symmetric and traceless,
\item Transverse: $X^A F_{A A_2 \cdots A_l} =0$\ .
\end{itemize}
These conditions are automatically imposed in the index-free notation that introduces an auxiliary vector $Z^A$ and turns the symmetric tensor into the polynomial,
\begin{align}
 F(X,Z) \equiv F_{A_1 \cdots A_l}(X)\,Z^{A_1} \cdots Z^{A_l}\ .
\end{align}
Imposing the traceless and transverse conditions is equivalent to restricting the polynomials on the subspace satisfying $Z^2 = 0$ and $Z\cdot X=0$, which leaves the gauge redundancy $F(X,Z + \lambda X) = F(X,Z) \, , \lambda \in \BR$ for the choice of the tensor field in the embedding space.

The index-free notation allows us to construct a gauge invariant for the symmetric traceless tensor $F(X,Z)$,
\begin{align}
	C_{ZX}^{AB} \equiv Z^A X^B - Z^B X^A \ ,
\end{align}
from which the two- and three-point functions are built out just by taking into account the conformal invariance.
For example, consider the two-point function of primary fields, $\CO_{\Delta,l}(X_i,Z_i)~(i=1,2)$, of spin $l$ and dimension $\Delta$.
It should be a degree $l$ gauge invariant function of both $Z_1$ and $Z_2$, but we can construct only one gauge invariant of degree one in $Z$'s,
\begin{align}
C_{Z_1 X_1}^{AB} C_{Z_2 X_2\,AB} 
		= 2[(Z_1 \cdot Z_2)(X_1 \cdot X_2)-(Z_1 \cdot X_2)(Z_2 \cdot X_1)]\ .
\end{align}
Hence the two-point function is determined by the homogeneity,
\begin{align}\label{2-pt}
	 \langle \CO_{\Delta,l}(X_1,Z_1)\,\CO_{\Delta,l}(X_2,Z_2)\rangle = \frac{1}{(X_{12})^{\Delta}} \left[\frac{(Z_1 \cdot Z_2)(X_1 \cdot X_2)-(Z_1 \cdot X_2)(Z_2 \cdot X_1)}{X_{12}} \right]^l\ .
\end{align}
Similarly, for the three-point function of two scalar primaries $\CO_{\Delta_i}(X_i)~(i=1,2)$ and a spin $l$ primary $\CO_{\Delta_3,l}(X_3,Z_3)$,
we have only one gauge invariant,
\begin{align}
	X_1\cdot C_{Z_3 X_3} \cdot X_2 = (Z_3\cdot X_1)(X_2\cdot X_3) - (Z_3\cdot X_2)(X_1\cdot X_3) \ ,
\end{align}
and the correlator is uniquely fixed by the conformal symmetry \cite{Costa:2011mg},
\begin{align}\label{3-pt}
\begin{aligned}
	\langle \CO_{\Delta_1} (X_1)\, &\CO_{\Delta_2}(X_2)\,\CO_{\Delta_3, l}(X_3, Z_3) \rangle \\
    	&= a_3\,  \frac{\left[(Z_3\cdot X_1)X_{23} - (Z_3\cdot X_2)X_{13}\right]^l}{(X_{12})^{(\Delta_1 + \Delta_2 - \Delta_3 + l)/2}\, (X_{23})^{(\Delta_2 + \Delta_3 - \Delta_1 + l)/2}\, (X_{31})^{(\Delta_3 + \Delta_1 - \Delta_2 + l)/2}} \ ,
\end{aligned}
\end{align}
up to a normalization factor $a_3$.

The index structure of correlators in the index-free notation can be recovered by acting the Todorov differential operators $D_A$ on $F(X,Z)$ \cite{Dobrev:1975ru}:
\begin{align}
D_A &= \left(\frac{d-2}{2} + Z \cdot \frac{\partial}{\partial Z} \right)\frac{\partial}{\partial Z^A}-\frac{1}{2}Z_A \frac{\partial^2}{\partial Z \cdot \partial Z} \ ,\\
F_{A_1 \cdots A_l}(X)  &= \frac{1}{l!(d/2 -1)_l}D_{A_1} \cdots D_{A_l}F(X,Z) \ ,
\end{align}
where $(a)_l = \Gamma(a+l)/\Gamma(a)$ is the Pochhammer symbol.
A spin $l$ symmetric traceless tensor field in the physical space $f_{a_1 \cdots a_l}(x)$ is obtained by pulling back the embedding tensor $F_{A_1 \cdots A_l}(X)$ on the Poincar{\'e} section,
\begin{align}
f_{a_1 \cdots a_l}(x) = \left.\frac{\partial X^{A_1}}{\partial x^{a_1}} \cdots \frac{\partial X^{A_l}}{\partial x^{a_l}} F_{A_1 \cdots A_l}(X) \right|_{X^A = (1,x^2,x^i)} \ .
\end{align}
The contraction of two symmetric traceless tensors $f_{a_1\cdots a_l}(x)$ and $g^{a_1\cdots a_l} (x)$ are given by the product of their encoding polynomials $F(X,Z)$ and $G(X,Z)$, replacing an auxiliary vector $Z$ with the Todorov operator $D$ for $F(X,Z)$ \cite{Costa:2011mg},
\begin{align}\label{Contracted_Tensors}
	f_{a_1\cdots a_l}(x)\, g^{a_1\cdots a_l} (x) = \frac{1}{l! (d/2 -1)_l} F(X, D)\, G(X, Z) \ .
\end{align}

\subsection{Conformal defects in embedding space}\label{ss:Defect_Embedding}

The full conformal symmetry $SO(d+1,1)$ is broken in the presence of a codimension-$m$ defect of a sphere or a planar type to the subgroup $SO(m) \times SO(d+1-m,1)$ that is the rotation group around and the conformal group on the defect.
To our best knowledge, conformal defects remain to be classified based on the residual symmetry unlike local operators living on them.
There appear to be, at least,  conformal defects in non-trivial representations under the $SO(m)$ symmetry, though we are not aware of any example.
In this paper, we restrict our attention to a scalar defect that 
has a simple realization in the embedding space \cite{Gadde:2016fbj,Billo:2016cpy}.
The embedding space formalism for conformal defects in non-trivial representations may be formulated in the same way as local operators in general tensor representations \cite{Costa:2014rya,Rejon-Barrera:2015bpa,Billo:2016cpy,Costa:2016hju}, but we postpone a detailed study to future work.

To specify the position of a codimension-$m$ conformal defect, we choose $m$ spacelike frame vectors $P_\alpha~(\alpha = 1, \cdots, m)$ and draw a hyperplane transverse to the vectors $P_\alpha$ and intersecting with the projective null cone,
\begin{align}\label{Defect_Def}
X^2 = 0 \ , \qquad P_{\alpha} \cdot X = 0\ . 
\end{align}
This configuration preserves the $SO(m) \times SO(d+1-m,1)$ symmetry that acts as the stabilizer subgroup of the $(d+2-m)$-dimensional hyperplane transverse to the $m$ frame vectors, and hence can be identified with the uplift of a codimension-$m$ defect into the embedding space.

Since there is a $GL(m)$ gauge-redundancy for the choice of the set of the frame vectors $P_{\alpha}$, we can gauge-fix them so that they are an orthonormal basis, $P_{\alpha} \cdot P_{\beta} = \delta_{\alpha \beta}$.
In this gauge, a defect of codimension-$m$ is fully characterized by the $m$-dimensional orthonormal frame and we denote it by $\CD^{(m)}(P_{\alpha})$.

Given a spherical or planer defect of codimension-$m$ in the physical space, one can uplift it to the embedding space by picking up any $(d+2-m)$ points on the defect, lifting them to the Poincar{\'e} section $X_{k}~ (k = 1, \cdots , d+2-m)$ and then solving $(d+2-m)\times m$ equations $X_{k} \cdot P_{\alpha} = 0$ for each $k$ and $\alpha$ to get a set of the frame vectors $P_{\alpha}$ \cite{Gadde:2016fbj}.
To illustrate how this procedure works, consider a codimension-$m$ spherical defect of radius $R$ centered at the origin and lying in a $(d+1-m)$-dimensional hyperplane spanned by the orthonormal basis $\vec e_j~ (j=1,\cdots d+1-m)$, each pointing to the $j^{th}$ coordinate in the physical space.
We can pick up $(d+2-m)$ points at $\vec{x}_j = R\,\vec{e}_j$ $(j=1,\cdots, d+1-m)$ and $\vec{x}_{d+2-m} = -R\,\vec{e}_1$ whose uplifts are at $X_j = (1,R^2,R\,\vec e_j)$ and $X_{d+2-m} = (1,R^2,-R\, \vec e_1)$ in the embedding space.
Solving the set of equations $X_k \cdot P_{\alpha}=0$ leads to a solution for the frame vectors,
\begin{align}
\begin{aligned}
P_{\alpha} &= (0,0,\vec e_{d+1-m+\alpha})\qquad \qquad (\alpha = 1,\cdots,m-1)\ ,\\
P_m &= \left(\frac{1}{R},-R,\vec 0 \right)\ .
\end{aligned}
\end{align}
When the center is shifted by $r$ in the $x_1$-direction, we can pick $(d+2-m)$ points at $X_j = \left(1,(R\, \vec e_j + r\, \vec e_1)^2,R\, \vec e_j + r\, \vec e_1 \right)$ and $X_{d+2-m}=\left(1,(R-r)^2,(r - R) \vec e_1 \right)$ on the defect in the embedding space.
In this case, the frame vectors are given by
\begin{align}\label{Sphere_Frame_Vectore}
\begin{aligned}
P_{\alpha} &= (0,0,\vec e_{d+1-m+\alpha})\qquad \qquad (\alpha = 1,\cdots,m-1)\ , \\
P_m &= \left(\frac{r}{R},-R+\frac{r^2}{R},\frac{r}{R} \vec e_1 \right)\ .
\end{aligned}
\end{align}

When acted on a codimension-$m$ defect described by the frame vectors, the generators of the conformal symmetry are represented by
\begin{align}
	L_{AB}(P_\alpha) \equiv \sum_{\alpha = 1}^m \left( P_{A\alpha} \frac{\partial}{\partial P^B_\alpha} - P_{B\alpha} \frac{\partial}{\partial P^A_\alpha}\right) \ .
\end{align}
The summation over $\alpha$ is taken to respect the $SO(m)$ symmetry under which the frame vectors $P_\alpha$ rotate in the subspace they span and thus the defect is invariant.

The center and radius of a spherical defect are read off from the frame vectors in the following way \cite{Gadde:2016fbj}.
We pick up a reference point $\Omega \equiv (0,1, 0^i)$ at infinity and construct a $GL(m)$ invariant null vector $C$ out of $P_\alpha$,
\begin{align}\label{Defect_Center}
	C = \frac{\Omega - 2(P^\alpha \cdot \Omega)P_\alpha}{4(P^\beta \cdot \Omega)(P_\beta \cdot \Omega)} \ .
\end{align}
Here the indices $\alpha$ of the frame vectors are raised and lowered with respect to the metric $\delta_{\alpha\beta}$.
This vector points to the center, and the radius $R$ is measured by the distance between the center and a point $X$ on the sphere, namely,
\begin{align}\label{Defect_Radius}
	R^2 = -2 C\cdot X = \frac{1}{4(P^\beta \cdot \Omega)(P_\beta \cdot \Omega)} \ .
\end{align}
It is easy to verify that the center \eqref{Defect_Center} and the radius \eqref{Defect_Radius} reproduce the correct values, $C = (1,r^2, r\,\vec{e_1})$ and $R$, for the spherical defect specified by the frame vectors \eqref{Sphere_Frame_Vectore}.

\subsection{Correlators in defect CFT}
In a defect CFT, the correlation functions of local operators are calculated in the presence of a defect operator,
\begin{align}
 \langle \CO(X_1) \cdots \CO(X_k) \rangle_{\CD} = \frac{1}{\langle \CD^{(m)}(P_\alpha)\rangle} \langle \CD^{(m)}(P_\alpha)\, \CO(X_1) \cdots \CO(X_k)  \rangle \ .
\end{align}
These types of correlators can be fixed in the embedding space by the symmetry and homogeneity in parallel with the correlators of local operators.
We will proceed our discussion with the normalization $\langle \CD^{(m)}(P_\alpha ) \rangle = 1$ in the following.

We consider as a simplest example the correlation function of a scalar defect and a spin $l$ operator,
\begin{align}
	\langle \CD^{(m)}(P_\alpha)\, \CO_{\Delta, l}(X, Z)\rangle \ .
\end{align}
This correlator should be a scalar function with correct dimensions in the index-free notation.
The scalar invariants we can construct out of the vectors $P_\alpha, X$ and $Z$ are $P_\alpha \cdot X$ and $P_\alpha\cdot Z$ (note $Z\cdot X=0$), and we must contract the index $\alpha$ to respect the $SO(m)$ symmetry.
Therefore the following three invariants are allowed to show up in the correlator,
\begin{align}
	(P^\alpha \cdot X)(P_\alpha \cdot X) \ , \qquad (P^\alpha \cdot Z)(P_\alpha \cdot Z) \ , \qquad (P^\alpha \cdot X)(P_\alpha \cdot Z) \ .
\end{align}
We further use the invariance under the shift $Z\to Z + c\, X$ that fixes the form of the correlator uniquely up to a factor \cite{Billo:2016cpy,Costa:2011mg,SimmonsDuffin:2012uy}:
\begin{align}\label{One_Point_Defect}
\begin{aligned}
	\langle \CD^{(m)}(P_\alpha)\, \CO_{\Delta, l}(X, Z)\rangle &= \frac{a_{\Delta, l}}{\left[ (P^\alpha \cdot X)(P_\alpha \cdot X)\right]^{(\Delta + l)/2}} \left[ (P^\beta \cdot C_{ZX} \cdot P^\gamma)(P_\beta \cdot C_{ZX} \cdot P_\gamma)\right]^{l/2}\ ,\\
    &= \frac{a_{\Delta, l}}{\left[ (P^\alpha \cdot X)(P_\alpha \cdot X)\right]^{(\Delta + l)/2}} \left[ (Z \cdot C_{P^\beta P^\gamma} \cdot X)(Z \cdot C_{P_\beta P_\gamma} \cdot X)\right]^{l/2}\ ,
\end{aligned}
\end{align}
where we used the relation $P^\beta \cdot C_{ZX} \cdot P^\gamma = Z \cdot C_{P^\beta P^\gamma} \cdot X$ in going from the first line to the second.
The correlators with non-zero spin $l>0$ vanish for $m=1$ as there is only one frame vector $P_1$ giving $C_{P_1 P_1} = 0$, which is consistent with the result in boundary CFT (see e.g.\,\cite{Liendo:2012hy}).
Note also that this correlator is parity invariant and makes sense only for even $l$ \cite{Billo:2016cpy}. 
Hence there are no non-vanishing parity invariant correlator for odd $l$.

One has to use the $SO(d + 1, 1)$-invariant $\epsilon$-tensor to get a non-vanishing parity odd correlator \cite{Costa:2011mg}.
Scalar invariants involving the $\epsilon$-tensor are build out by contracting the indices with the vectors $X$, $Z$ and the frame vectors $P_\alpha ~(\alpha = 1,\cdots, m)$ (or the dual frame vectors $\tilde P_{\tilde\alpha} ~(\tilde\alpha = 1,\cdots, d+2-m)$ introduced in the next subsection). 
Since each vector appears at most once due to the antisymmetry of the $\epsilon$-tensor parity odd invariants are possible to construct only for $m\le 2$.
There is one such an invariant for a codimension-two defect,
\begin{align}
	\epsilon_{A_1 \cdots A_{d} BC}\,  \tilde P_1 ^{A_1}\cdots  \tilde P_d^{A_d} X^B Z^C \ ,
\end{align}
and are $(d+1)$ invariants for a codimension-one defect,
\begin{align}
	\epsilon_{A_1 \cdots A_{d} BC}\,  \tilde P_{\tilde\alpha_1} ^{A_1}\cdots  \tilde P_{\tilde\alpha_d}^{A_d} X^B Z^C \ .
\end{align}
These are invariant under the shift $Z \to Z + c\,X$, and all parity odd  correlators for odd $l$ can be constructed by multiplying the parity odd invariants to the correlator \eqref{One_Point_Defect} with a spin $l-1$ primary operator.

The one-point functions of defect local operators $\hat \CO$ in symmetric traceless representations of both $SO(m)$ and $SO(d-m)$ are determined in a similar way to \eqref{One_Point_Defect} by introducing two auxiliary null vectors transverse and orthogonal to the defect that contract the $SO(d-m)$ and $SO(m)$ indices respectively \cite{Billo:2016cpy}.

\subsection{Dual frame and dual defect}\label{ss:duality}
We have seen a codimension-$m$ defect $\CD^{(m)}(P_\alpha)$ is completely characterized by a set of the frame vectors $P_\alpha$ as a hypersurface satisfying $P_\alpha\cdot X = 0$ $(\alpha = 1, \cdots, m)$ in the embedding space.
Instead of specifying the normal vectors one can fix the position of the same hypersurface in the dual frame spanned by the vectors $\tilde P_{\tilde \alpha}$ $(\tilde\alpha = 1, \cdots, d+2 - m)$ tangent to the hypersurface or equivalently transverse to the original frame \cite{Gadde:2016fbj},
\begin{align}
	P_\alpha \cdot \tilde P_{\tilde \alpha} = 0 \ .
\end{align}
In Euclidean CFT, the frame vectors are orthonormal, $P_\alpha\cdot P_\beta = \delta_{\alpha\beta}$, with respect to the flat metric $\delta_{\alpha \beta} = \text{diag}(1,1,\cdots,1)$, 
while the dual vectors are orthonormal, $\tilde P_{\tilde \alpha} \cdot \tilde P_{\tilde \beta} = \eta_{\tilde \alpha \tilde \beta}$, with respect to the Lorentzian metric $\eta_{\tilde \alpha \tilde \beta} = \text{diag}(1,\cdots,1, -1)$.

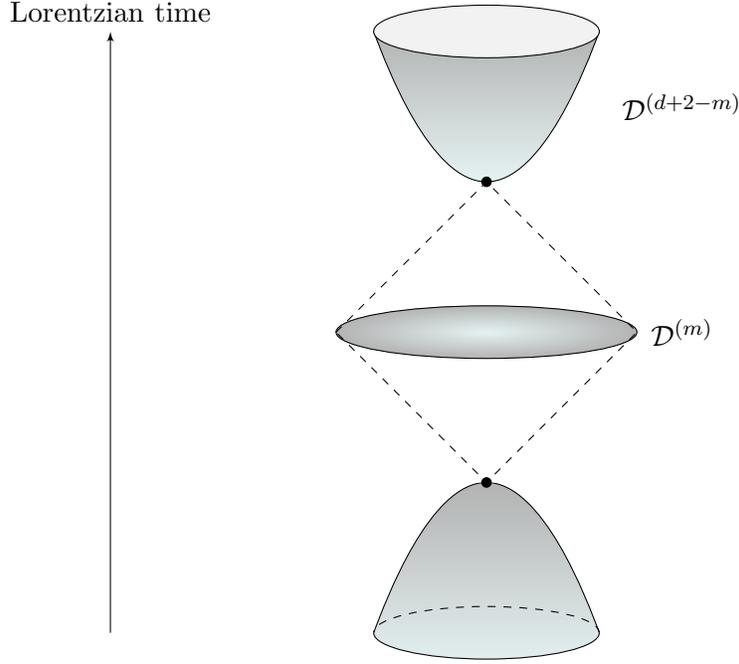
\begin{figure}[t]
\centering
\begin{tikzpicture}[>=latex']
    \shade[shading = radial, inner color = blue!50!green!10, outer color = gray!100!blue!60] (0,0) circle(2cm and 0.35cm);
    \draw (0,0) circle(2cm and 0.35cm);
    \draw[dashed] (-2,0) -- (0,2) -- (2,0) -- (0,-2) -- (-2,0);
    
    \shade[bottom color=blue!50!green!10, top color = gray!100!blue!60]
    (-1.5,4) parabola bend (0,2) (1.5,4);
    \draw (-1.5,4) parabola bend (0,2) (1.5,4);
    \draw[thin,fill=gray!10] (0,4) circle(1.5cm and 0.35cm);
    
    \begin{scope}
    \clip (-1.5, -2) -- ([shift=(180:1.5cm)]0,-4) arc (180:360:1.5cm and 0.35cm) -- (1.5, -2) --cycle;
    \shade[bottom color=blue!50!green!10, top color = gray!100!blue!60]
    (-1.5,-4.5) -- (-1.5,-4) parabola bend (0,-2) (1.5,-4) -- (1.5,-4.5);
    \draw (-1.5,-4) parabola bend (0,-2) (1.5,-4);
    \end{scope}
    
    \draw[dashed] ([shift=(0:1.5cm)]0,-4) arc (0:180:1.5cm and 0.35cm);
    \draw ([shift=(180:1.5cm)]0,-4) arc (180:360:1.5cm and 0.35cm);
    
    \fill (0,2) circle (2pt);
    \fill (0,-2) circle (2pt);
    
    \draw[->] (-5, -4) -- (-5, 4) node[above] {Lorentzian time};
    \node at (2.6,0) {$\CD^{(m)}$};
    \node at (2.6,3) {$\CD^{(d+2-m)}$};
\end{tikzpicture}
\caption{A codimension-$m$ spherical defect and its dual defect}
\label{fig:general-duality}
\end{figure}

By exchanging the roles of the frame and dual frame vectors, we can define a codimension-$(d+2-m)$ defect $\CD^{(d+2-m)}(\tilde P_{\tilde\alpha})$ as a hypersurface normal to the dual frame vectors $\tilde P_{\tilde\alpha}$ of $\CD^{(m)}(P_\alpha)$.
We call this object $\CD^{(d+2-m)}(\tilde P_{\tilde\alpha})$ a \emph{dual defect} of $\CD^{(m)}(P_\alpha)$.
It is clear from this definition that the dual of the dual defect $\CD^{(d+2-m)}(\tilde P_{\tilde\alpha})$ returns to the original defect $\CD^{(m)}(P_\alpha)$.
Hence we obtain an intriguing duality between a codimension-$m$ defect and a codimension-$(d+2-m)$ defect:
\begin{align}\label{Duality}
	\text{codimension}:\qquad m \quad \longleftrightarrow \quad d+2-m \ .
\end{align}
See figure \ref{fig:general-duality} for the illustration.
A correlator involving a defect $\CD^{(m)}(P_\alpha)$ is a function of the frame vectors $P_\alpha$, but it can be equally described as a function of the dual frame vectors $\tilde P_{\tilde\alpha}$.
To fix the normalization of the dual defect we further require the invariance of the defect correlator under the duality,
\begin{align}
	\langle \CD^{(m)}(P_\alpha) \cdots \rangle = \langle \CD^{(d+2-m)}(\tilde P_{\tilde\alpha}) \cdots \rangle \ .
\end{align}
Then we can freely replace a defect with its dual defect inside any correlator.

Now we want to determine the position of the dual defect $\CD^{(d+2-m)}(\tilde P_{\tilde\alpha})$ from the data of the original defect $\CD^{(m)}(P_\alpha)$.
Namely we will represent the center $\tilde C$ and radius $\tilde R$ of the dual defect defined by
\begin{align}
	\tilde C = \frac{\Omega - 2(\tilde P^{\tilde\alpha} \cdot \Omega)\tilde P_{\tilde\alpha}}{4(
    \tilde P^{\tilde\beta} \cdot \Omega)(\tilde P_{\tilde\beta} \cdot \Omega)} \ , \qquad \tilde R^2 = \frac{1}{4(
    \tilde P^{\tilde\beta} \cdot \Omega)(\tilde P_{\tilde\beta} \cdot \Omega)} \ ,
\end{align}
in terms of the original ones $C$ and $R$.
To this end, we first calculate the distance $r$ between the two centers $C$ and $\tilde C$, both of which are null vectors,
\begin{align}\label{Center_distance}
\begin{aligned}
	r^2 &= -2 C\cdot \tilde C \ ,\\
    	&= \frac{(P^\alpha \cdot \Omega)(P_\alpha \cdot \Omega) + (
    \tilde P^{\tilde\alpha} \cdot \Omega)(\tilde P_{\tilde\alpha} \cdot \Omega) }{4(P^\beta \cdot \Omega)(P_\beta \cdot \Omega) (
    \tilde P^{\tilde\beta} \cdot \Omega)(\tilde P_{\tilde\beta} \cdot \Omega) }\ ,\\
    &=R^2 + \tilde R^2 \ .
\end{aligned}
\end{align}
The distance $r$ looks non-vanishing at first sight, but one can show the two centers coincide, $r=0$, as follow.
Recall the frame and dual frame vectors span the orthonormal basis in the embedding space,
\begin{align}
 	P^{\,\alpha A}\, P_\alpha^B + \tilde P^{\,\tilde \alpha A}\, \tilde P_{\tilde\alpha}^B = \eta^{AB} \ .
\end{align}
Multiplying the reference vector $\Omega$ twice to contract the indices of this expression, one ends up with the identity,
\begin{align}
	(P^\alpha \cdot \Omega)(P_\alpha \cdot \Omega) + (
    \tilde P^{\tilde\alpha} \cdot \Omega)(\tilde P_{\tilde\alpha} \cdot \Omega) = 0 \ ,
\end{align}
proving our statement.
This result implies $\tilde R^2 = - R^2$ from \eqref{Center_distance}, and one may wonder how it is possible to draw a hypersphere of an imaginary radius in the embedding space.
This is simply due to the fact that one of the dual frame vectors is not spacelike.
Thus the defining equations $\tilde X\cdot \tilde P_{\tilde\alpha} = 0$ for the dual defect do not have any solution in Euclidean CFT.
Instead they are able to have a unique solution in Lorentzian CFT.

The following example illustrates how the duality works in practice.
Consider the dual of a codimension-$m$ spherical defect whose frame vectors are given by \eqref{Sphere_Frame_Vectore}.
In the dual frame, the frame vectors are chosen to be
\begin{align}
\begin{aligned}
	\tilde P_{\tilde \alpha} &= (0,0, \vec e_{\alpha}) \qquad\qquad (\alpha = 1,\cdots, d+1-m) \ , \\
    \tilde P_{d+2-m} &= (1/R, R, \vec 0) \ .
\end{aligned}
\end{align}
Note that these vectors are spacelike except the last one.
Let us see the configuration of the dual defect by regarding $\tilde P_{\tilde\alpha}$ as the frame vector of a codimension-$(d+2-m)$ defect .
In the embedding space, the coordinate is given by the null vector $\tilde X = (1, \tilde x^2, \vec{\tilde x})$.
The position of the dual defect is fixed by specifying $m$ points $\tilde X_{k=1,\cdots, m}$ satisfying the conditions $\tilde X_k \cdot \tilde P_{\tilde\alpha} = 0$.
Namely we must solve the following set of equations,
\begin{align}
	\vec {\tilde x} \cdot \vec e_\alpha = 0 \quad (\alpha = 1,\cdots, d+1-m) \ , \qquad \tilde x^2 = - R^2 \ .
\end{align}
As a particular solution, we can find
\begin{align}
	\tilde X_j^{(\pm)} = (1, - R^2, \ell\,\vec e_{d+1-m+j} \pm i \ell'\,\vec e_d) \quad (j=1,\cdots, m-1) \ , \qquad \tilde X_m^{(\pm)} = (1, - R^2, \pm i R\,\vec e_{d})\ ,
\end{align}
where $\ell$ is a non-negative parameter and we define $\ell'$ by
\begin{align}
	\ell' = \sqrt{\ell^2 + R^2} \ .
\end{align}
All of these points lie on the imaginary Euclidean time, or equivalently on the real Lorentzian time $\tilde x_0 \equiv i \tilde x_d$.
In the Lorentzian signature, the dual defect is a $(m-2)$-dimensional hyperbolic surface given by
\begin{align}\label{Dual_position}
	- \tilde x_0^2 + \sum_{i=1}^{d-1} \tilde x_i^2 = - R^2 \ ,\qquad \tilde x_i = 0 \quad (i=1,\cdots d+1-m) \ ,
\end{align}
in $\BR^{d-1,1}$, and the two sets of points $\tilde X_k^{(+)}$ and $\tilde X_k^{(-)}$ are located on one of the two branches of the hyperbolic surface (see figure \ref{fig:general-duality}).

\subsection{Codimension-two defects}\label{ss:Codim-two}
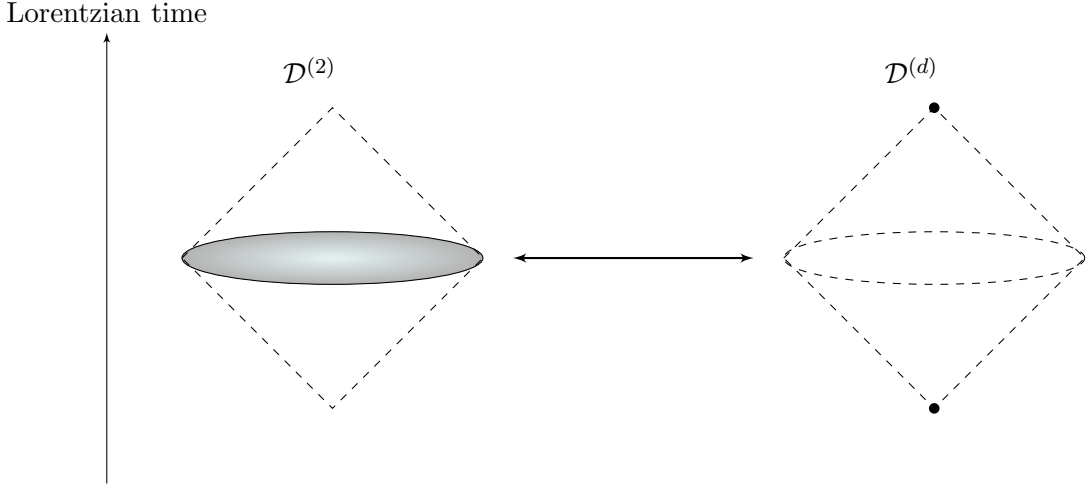
\begin{figure}[t]
\centering
\begin{tikzpicture}[>=latex']
    \shade[shading = radial, inner color = blue!50!green!10, outer color = gray!100!blue!60] (-4,0) circle (2cm and 0.35cm);
    \draw (-4,0) circle (2cm and 0.35cm);
    \draw[dashed] (-6,0) -- (-4,2) -- (-2,0) -- (-4,-2) -- (-6,0);
    
    \draw[dashed] (4,0) circle(2cm and 0.35cm);
    \draw[dashed] (2,0) -- (4,2) -- (6,0) -- (4,-2) -- (2,0);
    
    \fill (4,2) circle (2pt);
    \fill (4,-2) circle (2pt);
    
    \draw[->] (-7, -3) -- (-7, 3) node[above] {Lorentzian time};
    \draw[<->,thick] (-1.6,0) -- (1.6,0);
    
    \node at (-4.3,2.5) {$\CD^{(2)}$};
    \node at (3.7,2.5) {$\CD^{(d)}$};
\end{tikzpicture}
\caption{The defect duality between a codimension-two defect and a codimension-$d$ defect when $d=3$}
\label{fig:duality}
\end{figure}

We treat a codimension-two defect separately as a special case of the dual defect being local (= codimension-$d$).
In this case, the position of the dual defect given by the condition \eqref{Dual_position} consists of a pair of points in Lorentzian signature,
\begin{align}
	\tilde x_i = 0 \quad (i=1,\cdots, d-1) \ , \qquad \tilde x_0 = \pm R \ .
\end{align}
These are actually at the tips of the causal diamond of the original spherical defect.
Thus we can associate a pair of local operators to a given codimension-two defect (see figure \ref{fig:duality}).

For a more general configuration, we can determine the position $\tilde X$ of the dual defect as follows.
We can expand the coordinates $\tilde X$  by the frame vectors $P_\alpha$ as
\begin{align}
	\tilde X = a P_1 + b P_2 \ ,
\end{align}
as the dual defect lives in the normal plane to the defect.
This vector automatically satisfies the condition $\tilde X\cdot \tilde P_{\tilde \alpha} = 0$.
For being a null vector in the embedding space, $\tilde X^2 = 0$, the position vector has to take the form,
\begin{align}\label{FrameToDualCoord}
	\tilde X_1 = a_1(P_1 + i P_2) \ ,\qquad \tilde X_2 = a_2(P_1 - i P_2) \ ,
\end{align}
with undetermined constants $a_1$ and $a_2$ that can be fixed by the condition $(\tilde X)^+ = 1$.
As a result, the position $X$ of the codimension-two defect has to be transverse to the dual defect due to the relation $X\cdot P_\alpha = 0$:
\begin{align}\label{CausalDualRelation}
	X \cdot \tilde X_k = 0 \qquad (k=1,2)\ .
\end{align}

This relation was observed by \cite{deBoer:2016pqk,Czech:2016xec} in analyzing the kinematic space of a spherical entangling surface, where
they characterized a spherical defect by the two points at the tips of the causal diamond.
Namely for a given pair of timelike separated points $\tilde x_1^\mu$ and $\tilde x_2^\mu$ in flat Lorentzian space, one can define a codimension-two spherical surface as the intersection of the past and future light cones.
Denoting the coordinates of the spacetime by $x^\mu$, the position of the spherical surface is given by
\begin{align}
	(x- \tilde x_1)^2 = 0 \ , \qquad (x-\tilde x_2)^2 = 0 \ .
\end{align}
In the embedding space, these conditions lift to
\begin{align}
	X \cdot \tilde X_1 = 0 \ , \qquad X\cdot \tilde X_2 = 0 \ .
\end{align}
These are the same as \eqref{CausalDualRelation} and we have shown the equivalence of our construction of the dual defects for $m=2$ and the causal diamond construction of a codimension-two spherical surface by \cite{deBoer:2016pqk,Czech:2016xec}.

We can rewrite the codimension-two defect correlator as a function of the dual positions $\tilde X_1, \tilde X_2$ through the relation \eqref{FrameToDualCoord}.
Since the index $\alpha$ is always contracted inside correlators, we find it convenient to employ the following relation,
\begin{align}
	\sum_{\alpha = 1,2} P_\alpha^A P_\alpha^B = -2\, \frac{\tilde X_1^A \tilde X_2^B + \tilde X_1^B \tilde X_2^A}{\tilde X_{12}} \ .
\end{align}
Note that the left hand side is the projection operator to the hyperplane normal to the defect.
Given this relation, the defect one-point function \eqref{One_Point_Defect} of a spin $l$ operator becomes
\begin{align}
\begin{aligned}
	\langle \CD^{(2)}(P_\alpha)\, &\CO_{\Delta, l}(X_3, Z_3)\rangle \\
    &= a_{\Delta, l}\, (-1)^{\Delta/2}\,2^{l/2}\, \left[ \frac{\tilde X_{12}}{(\tilde X_{1}\cdot X_{3}) (\tilde X_{2}\cdot X_{3}) }\right]^{(\Delta+l)/2}  \left[ \frac{(Z_3 \cdot \tilde X_2)(\tilde X_{1}\cdot X_{3}) -(Z_3 \cdot \tilde X_1)(\tilde X_{2}\cdot X_{3}) }{\tilde X_{12}}\right]^{l}\ .
\end{aligned}
\end{align}
The right hand side is, up to the normalization, precisely the scalar-scalar-spin $l$ three-point function \eqref{3-pt} with conformal dimensions $\Delta_1 = \Delta_2 = 0$ and $\Delta_3 = \Delta$.
This result is consistent with our proposal of the duality \eqref{Duality} and we can identify the codimension-$d$ dual defect with a pair local operators sitting on the dual positions $\tilde X_1$ and $\tilde X_2$,
\begin{align}\label{Dual_Defect_m=2}
	\CD^{(d)}(\tilde P_{\tilde \alpha}) = \Phi(\tilde X_1)\,\Phi(\tilde X_2) \ .
\end{align}

After introducing the defect OPE blocks in the next section, we revisit the codimension-two case in section \ref{ss:OPE_block} where the duality of defects connects our results of codimension-two with the OPE blocks of local operators examined by \cite{Czech:2016xec,deBoer:2016pqk}.

\section{Defect OPE blocks}\label{ss:Defect_OPE_block}

\begin{figure}[t]
\centering
\begin{tikzpicture}
	\shade[ball color=blue!30!green!20, opacity=0.70] (0,0) circle (1cm);
    \draw[dashed] (0,0) circle (2cm);
    \node[fill=white] at (1.2, 1.3) {$\CD^{(m)}$};
    
    \node at (3,0) {\Large $=$};
    
    \node at (5,0) {\Large $\sum_n c_{\CO_n}^{(m)} \,R^{\Delta_n}$};
    \node at (6.7,-0.5) {\Large $\CO_n$};
    \fill (6.7,0) circle (2pt);
    \node at (8,0) {\Large $+ \quad \cdots$};
\end{tikzpicture}
\caption{A schematic picture of the OPE for conformal defects}
\label{fig:Defect-Exp}
\end{figure}
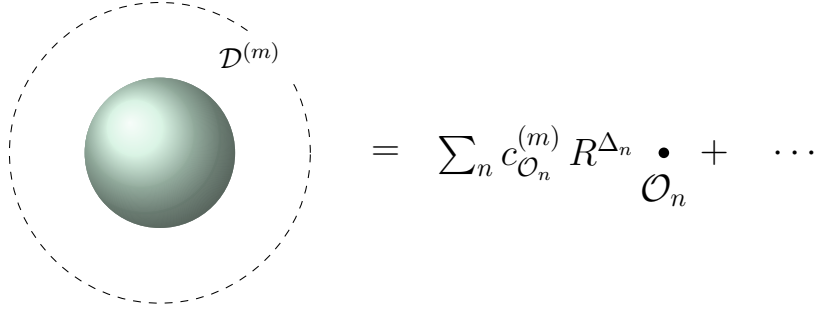

A defect operator is characterized by specifying a corresponding state on a surface large enough to enclose the defect.
Invoking the state-operator correspondence, any defect admits the OPE by a set of bulk local operators located at the center \cite{Berenstein:1998ij,Gomis:2009xg,Gadde:2016fbj} (see figure \ref{fig:Defect-Exp}),
\begin{align}\label{DefectOPE}
	\CD^{(m)}(P_\alpha) = \langle \CD^{(m)}(P_\alpha) \rangle \left[ \sum_{n} c_{\CO_n}^{(m)} \, R^{\Delta_n}\, \CO_{n}(C) + (\text{descendants})\right] \ ,
\end{align}
where $\CO_{n}(C)$ are a set of primary operators at the center $C$ of the spherical defect of radius $R$ fixed by the normal vectors $P_\alpha$ as in \eqref{Defect_Radius}.
The coefficients $c_{\CO_n}^{(m)}$ compensate the tensorial structure to make the term in the square bracket be a scalar.
We will present the structure of the defect OPE in a compact form by employing the shadow formalism that allows to collect the contributions from a primary operator $\CO_{n}$ and its descendants into a block, a non-local function of $\CO_{n}$.

\subsection{Integral representation and monodromy prescription}
\label{ss:IntegralRep}

The descendant terms in the defect OPE \eqref{DefectOPE} are, in principle, fully determined by the conformal symmetry so as to reproduce the correlation function of the defect and a local operator.
The defect OPE can be decomposed into the contributions from each primary operator $\CO_{n}$ and their descendant terms that transform in the same way as $\CO_n$ under the conformal group, 
thus it will be convenient to introduce the \emph{defect OPE block} $\CB^{(m)}[P_\alpha, \CO_{n}]$ that packages the contribution from the conformal multiplet of $\CO_{n}$,
and represent the defect OPE in the block decomposed form,
\begin{align}
	\CD^{(m)}(P_\alpha) = 
    \sum_n 
    \,\CB^{(m)}[P_\alpha,\CO_{n}]  \ .
\end{align}
The defect OPE block $\CB^{(m)}[P_\alpha, \CO_{n}]$ has to transform in the same way as the defect operator under the conformal transformation, and should be fixed by requiring it reproduces the same correlator with any local operator as the defect does,
\begin{align}\label{Block_Defining}
	\langle \CB^{(m)}[P_\alpha, \CO_{n}]\, \CO_{n} (X)\rangle = \langle \CD^{(m)}(P_\alpha)\,\CO_{n} (X)\rangle \ .
\end{align}
In order to calculate the defect OPE block, one may assume the block is given by the form,
\begin{align}
	\CB^{(m)}[P_\alpha, \CO_{n}] = f_{\Delta} (P_\alpha, C, \partial_C)\,\CO_{n}(C) \ ,
\end{align}
with a polynomial $f_{\Delta_n}(P_\alpha, C, \partial_C)$ of a differential operator, and fix the polynomial order by order by comparing the both hand sides of the defining relation \eqref{Block_Defining}.
This approach is systematic enough to solve $f_{\Delta_n}(P_\alpha, C, \partial_C)$ perturbatively, but not so efficient to work out if one wants a closed expression of the defect OPE block.

Instead of solving the relation \eqref{Block_Defining} directly, we present a transparent derivation of an integral representation of the block using the spectral decomposition of the identity operator $\bf 1$ by the irreducible representations of the conformal group \cite{Ferrara:1972ay,Ferrara:1973vz,Ferrara:1972xe,Ferrara:1972uq,Mack:1975jr,Mack:1976pa,Fradkin:1996is},
\begin{align}\label{Identity}
	{\bf 1} = \sum_n\, |\CO_n |\ .
\end{align}
The projector $|\CO_n|$ is the projection operator onto the conformal multiplet of the primary operator $\CO_n$.
For bosonic operators, $\CO_n$ are labeled by the conformal dimension $\Delta$ and the Young diagram $\bf{Y}$ specifying the irreducible representation $n = (\Delta, \bf{Y})$ of the rotation group $SO(d)$.
Hereafter, we focus on the spin $l$ case for the sake of simplicity where the projector takes the following form \cite{SimmonsDuffin:2012uy},
\begin{align}\label{Projector}
	|\CO_{\Delta, l }| = \frac{1}{\CN_{\Delta, l }}\,\int D^d X\, |\CO_{\Delta, l} (X, D_Z)\rangle \, \langle \tilde \CO_{d-\Delta, l} (X, Z) | \ .
\end{align}
The shadow field $\tilde \CO_{d-\Delta, l }$ in the integrand is a non-local operator build from the primary field $\CO_{\Delta, l}$ by the integral shadow transform,
\begin{align}\label{Shadow_Def}
		\tilde\CO_{d-\Delta,l}(X, Z) \equiv \frac{1}{\CN_{d- \Delta, l }} \int D^d Y\, \frac{1}{(-2X\cdot Y)^{d-\Delta + l}}\,\CO_{\Delta,l}(Y, -2\, C_{ZX}\cdot Y) \ .
\end{align}
The normalization constant $\CN_{\Delta, l}$ is fixed to be
\begin{align}
	\CN_{\Delta, l } = \pi^{d/2} \,(d-\Delta - 1)_l\,\frac{\Gamma (d/2 - \Delta)}{\Gamma(\Delta + l)} \ ,
\end{align}
which assures performing the shadow transformation twice gets back to the original field, $\tilde{\tilde{\CO}}_{\Delta, l} = \CO_{\Delta, l}$ \cite{Dolan:2011dv,Sleight:2017fpc}.
The shadow field transforms in the same representation of the conformal group as a primary field $\CO_{d-\Delta, l}$, which guarantees the projector \eqref{Projector} is a conformal scalar of dimension zero.
The shadow primaries $\tilde\CO_{d-\Delta,l}$ also span the complete basis, but they are orthogonal to the original primary basis \cite{Ferrara:1972uq},
\begin{align}\label{Shadow_2pt}
	\langle \tilde\CO_{d-\Delta,l}(X, Z) \, \CO_{\Delta,l}(X', Z') \rangle = \frac{\CN_{\Delta,l}}{l! (d/2 -1)_l}\,\delta^{d+2} (X - X')\, (Z\cdot Z')^l\ .
\end{align}
Here we defined the delta function in the embedding space such that
\begin{align}
	\int D^d X \, f(X)\,\delta^{d+2}(X - Y) = f(Y) \ ,
\end{align}
and the normalization constant in \eqref{Shadow_2pt} is fixed by requiring the projector being trivial in the correlator, $\langle |\CO_{\Delta, l }| \,\CO_{\Delta,l}(X,Z)\,\cdots\rangle = \langle \CO_{\Delta,l}(X,Z)\cdots \rangle.$

The projectors in representations other than the symmetric traceless ones are constructed for general tensors in any dimensions \cite{Costa:2014rya,Rejon-Barrera:2015bpa,Costa:2016hju} and for spinors in $d=4$ dimensions \cite{SimmonsDuffin:2012uy}.
While we can determine the blocks in any representation explicitly with a slight complication, we will focus on the contributions from spin $l$ primary operators to make things clear.

Inserting the spectral decomposition of the identity \eqref{Identity} into a correlator, we can expand the defect operator into a sum of the irreducible representations of the conformal group,
\begin{align}
\begin{aligned}
	\langle \CD^{(m)}(P_\alpha)\, \cdots \rangle &= \sum_{\Delta, l}\,\langle \CD^{(m)}(P_\alpha) |\CO_{\Delta, l }|  \cdots \rangle + (\text{other irrep.})\ , \\
    	&= \sum_{\Delta, l}\,\frac{1}{\CN_{\Delta, l }}\int D^d X\, \langle \CD^{(m)}(P_\alpha) \,\CO_{\Delta, l} (X, D_Z)\rangle \, \langle \tilde \CO_{d-\Delta, l} (X, Z)\, \cdots \rangle + (\text{other irrep.}) \ ,\\
        &= \sum_{\Delta, l}\,\frac{1}{\CN_{\Delta, l }}\int D^d X\, \langle \tilde \CO_{d-\Delta, l} (X, D_Z)\, \cdots \rangle \,\langle \CD^{(m)}(P_\alpha) \,\CO_{\Delta, l} (X, Z)\rangle + (\text{other irrep.}) \ ,
\end{aligned}
\end{align}
where we switched the order of the two correlators in the integrand with the exchange of the Todorov operator $D_Z$ and the auxiliary vector $Z$ in the third equality.
Since this relation has to hold for any correlator we conclude that the defect OPE block in the spin $l$ representation takes the following form:
\begin{align}\label{Defect_OPE_Block_candidate}
	\CB^{(m)}[P_\alpha, \CO_{\Delta,l}] = \frac{1}{\CN_{\Delta, l }}\int D^d X \, \tilde\CO_{d-\Delta, l} (X,D_Z)\,\langle \CD^{(m)}( P_\alpha) \,\CO_{\Delta,l}(X,Z)\rangle \ .
\end{align}
The blocks in the other irreducible representations are given in a similar manner by using the shadow formalism as well.
Substituting the correlator \eqref{One_Point_Defect} into \eqref{Defect_OPE_Block_candidate} provides us the concrete integral form of the defect OPE block.

This is almost what we want, but not exactly the true defect OPE block.
The expression (\ref{Defect_OPE_Block_candidate}) is invariant under flipping the roles of $\CO$ and $\tilde{\CO}$ and contains the \emph{shadow block} in addition to the true one.
To illustrate it consider the following function of a scalar primary $\CO_{\Delta}$,\footnote{In this paper, all symbols labeled by only conformal dimension $\Delta$ are defined for scalar operators.}
\begin{align}
F[P_\alpha] =  \int D^d X D^d Y \,\langle \CD^{(m)}(P_{\alpha}) \,\CO_{\Delta}(X)\rangle\, \frac{1}{(-2X \cdot Y)^{d-\Delta}} \,\CO_{\Delta}(Y)\ .
\end{align}
If we integrate over $Y$ first, we recover, up to a normalization constant, the candidate block \eqref{Defect_OPE_Block_candidate} with $l=0$ as a result of the definition of the shadow operator \eqref{Shadow_Def}.
Alternatively we can integrate over $X$ first and arrive at the different form,
\begin{align}
F[P_\alpha] = \int D^d Y \, \CO_{\Delta} (Y)\,\langle \CD^{(m)}( P_\alpha) \,\tilde\CO_{d-\Delta}(Y)\rangle \ ,
\label{Defect_Shadow_Block}
\end{align}
Hence we have shown that the function $F[P_\alpha]$ is invariant under the flip of the primary $\CO_\Delta$ and its shadow $\tilde\CO_{d-\Delta}$.
Repeating the same argument for higher spin fields yields the flipped representation of the candidate block \eqref{Defect_OPE_Block_candidate},
\begin{align}\label{Defect_OPE_Block_flipped}
		\CB^{(m)}[P_\alpha, \CO_{\Delta,l}] = \frac{1}{\CN_{d-\Delta, l }}\int D^d X \, \CO_{\Delta, l} (X,D_Z)\,\langle \CD^{(m)}( P_\alpha) \,\tilde \CO_{d-\Delta,l}(X,Z)\rangle \ ,
\end{align}
The equivalence between the two expressions \eqref{Defect_OPE_Block_candidate} and \eqref{Defect_OPE_Block_flipped} originates from the fact that the identity operator \eqref{Identity} also has the spectral decomposition by the shadow projector $|\tilde \CO_{\Delta,l}|$ as well.
Denoting the true and shadow blocks by $g_\CO$ and $g_{\tilde\CO}$ respectively, the candidate defect OPE block \eqref{Defect_OPE_Block_candidate} must be their linear combination,
\begin{align}
\CB^{(m)}[P_\alpha, \CO_{\Delta}] = g_{\CO} + K\, g_{\tilde\CO} \ ,
\end{align}
where $K$ is some constant we are not interested in.

To pick up from \eqref{Defect_OPE_Block_candidate} the true block $g_\CO$ that behaves desirably in small radius limit, we have to remove the shadow contribution $g_{\tilde\CO}$.
Here we adopt the monodromy prescription that was originally developed by \cite{SimmonsDuffin:2012uy} for extracting the conformal block of local operators in the shadow formalism. 
In our case, we consider the monodromy $M: P_\alpha \rightarrow e^{-2\pi i} P_\alpha$ rotating the phases of all the frame vectors simultaneously. 
Such a monodromy operator $M$ is generated by $M = \exp \left( -2\pi i \sum_\alpha P_\alpha \cdot \partial/\partial P_\alpha\right)$ for the frame vector $P_\alpha$.
We argue this operator acts on $g_{\CO}$ and $g_{\tilde\CO}$ as follows:
\begin{align}
\begin{aligned}
g_{\CO} &\quad \rightarrow \quad e^{2 \pi i \Delta}\,g_{\CO} \ ,\\
g_{\tilde\CO} & \quad \rightarrow \quad e^{2\pi i (d-\Delta)}\,g_{\tilde\CO} \ .
\end{aligned}
\end{align}
Hence projecting \eqref{Defect_OPE_Block_candidate} to the appropriate eigenspace of $M$ gets rid of the shadow block $g_{\tilde\CO}$,
\begin{align}\label{Defect_OPE_Block}
\CB^{(m)}[P_\alpha, \CO_{\Delta,l}]= \left. \frac{1}{\CN_{\Delta, l }} \int D^d X \, \tilde\CO_{d-\Delta,l} (X,D_Z)\,\langle \CD^{(m)}( P_\alpha) \,\CO_{\Delta,l}(X,Z)\rangle \right|_{M=e^{2 \pi i \Delta}}.
\end{align}
This is one of our main results advocated in Introduction, and 
we will discuss the implications of this expression for the rest of this paper.

Now we show the monodromy prescription reproduces the correct limiting behavior \eqref{DefectOPE} in the small radius limit, $R\to 0$.
We first represent the limit using the frame vectors with the relation \eqref{Defect_Radius}.
Without loss of generality, we can choose the frame vectors so that only one of them has the $+$ components
\begin{align}
\begin{aligned}
	P_{\alpha} &= (0,0,P_\alpha^i)&\qquad & (\alpha = 1, \cdots, m-1) \ , \\
    P_m &= (p^+, p^-, p^i)& \qquad & (p^+ > 0)\ .
\end{aligned}
\end{align}
With these choices, the center \eqref{Defect_Center} and radius \eqref{Defect_Radius} of the spherical defect are determined by the last vector $P_m$:
\begin{align}
	C = \frac{\Omega + p^+\, P_m}{(p^+)^2}\ , \qquad\qquad R = \frac{1}{p^+} \ .
\end{align}
Thus the small radius limit is equivalent to the large $p^+$ limit,
\begin{align}
	R \to 0 \qquad \longleftrightarrow\qquad p^+ \to \infty \ ,
\end{align}
while keeping the other frame vectors finite.
In taking the large $p^+$ limit, it will be convenient to introduce a rescaled vector $Q\equiv P_m/p^+$ with the norm $Q^2 = 1/(p^+)^2$.
This vector approaches to the center in the limit,
\begin{align}\label{Center_Q}
	Q = C + O(R^2)\ , \qquad \qquad R\to 0 \ .
\end{align}

Next we proceed to take the small radius limit of the integral \eqref{Defect_OPE_Block_candidate}.
To simplify the presentation, we concentrate on the scalar case ($l=0$).
Recall that the correlator of the defect and local operator inside the integrand takes the form \eqref{One_Point_Defect}.
It is a function of the invariant $(P^\alpha\cdot X)(P_\alpha\cdot X)$ for $l=0$ in which the $\alpha = m$ term of order $O\left((p^+)^2\right)$ dominates while the others are of order $O(1)$.
Ignoring an overall factor and keeping track of the $R$ and coordinate dependences, the leading part of the correlator is
\begin{align}\label{One_Point_Small}
	\langle \CD^{(m)}( P_\alpha) \,\CO_{\Delta}(X)\rangle \sim \frac{R^\Delta}{(Q\cdot X)^\Delta} + \cdots \ ,
\end{align}
in the small radius limit.
The relation \eqref{Center_Q} allows us to replace $Q$ with the center $C$ in $R\to 0$.
Finally substituting the expansion \eqref{One_Point_Small} into the integral expression \eqref{Defect_OPE_Block_candidate} leads to
\begin{align}
\begin{aligned}
	\CB^{(m)}[P_\alpha, \CO_{\Delta}] &\sim R^\Delta \int D^d X \, \frac{1}{(C\cdot X)^\Delta}\,\tilde\CO_{d-\Delta} (X) + \cdots \ ,\\
    &\sim R^\Delta\, \CO_{\Delta}(C) + \cdots \ .
\end{aligned}
\end{align}
This is what we want in the small radius limit.
Naively we do not need the monodromy prescription to project out the shadow block.
If, however, we started with the flipped form \eqref{Defect_OPE_Block_flipped} we would have ended up with the different boundary condition $\CB^{(m)} \sim R^{d-\Delta}\, \tilde \CO_{d-\Delta}$ in the small $R$ limit.
It implies that the limiting behavior of the candidate block has two leading terms,
\begin{align}
\begin{aligned}
	\CB^{(m)}[P_\alpha, \CO_{\Delta}] &\sim R^\Delta\, \CO_{\Delta}(C) + \cdots \\
    	&\qquad + R^{d-\Delta}\, \tilde \CO_{d-\Delta}(C) + \cdots \ ,
\end{aligned}
\end{align}
which is manifestly invariant under the exchanges $\Delta \leftrightarrow d-\Delta$ and $\CO \leftrightarrow \tilde\CO$.
We therefore must impose the monodromy prescription on the block \eqref{Defect_OPE_Block} to obtain the correct limiting behavior in the small radius expansion.

A similar argument for the $l>0$ case fixes the asymptotic behavior of the defect OPE as well (see appendix \ref{ap:SmallRadius} for details).
We will show our defect OPE block \eqref{Defect_OPE_Block} reproduces the OPE block defined by \cite{Czech:2016xec,deBoer:2016pqk} when $m=2$ in section \ref{ss:OPE_block}, which also supports the validity of our prescription.

\subsection{Constraint equations}
We have given the integral representation of the defect OPE blocks \eqref{Defect_OPE_Block} in the previous section, 
and here we show they obey a set of constraint equations characterizing their features.

The first constraint is the conformal Casimir equation that follows from the fact that the blocks are in the irreducible representation of the conformal group, 
\begin{align}\label{Defect_OPE_Eigen}
	\left(L^2(P_\alpha) + \CC_{\Delta, l}\right) \CB^{(m)}[P_\alpha, \CO_{\Delta, l}] = 0 \ , \qquad \CC_{\Delta, l} = \Delta (\Delta -d) + l(l+d-2)\ ,
\end{align}
where $L^2(P_\alpha)$ is the quadratic Casimir operator represented in the frame vectors
\begin{align}
	L^2(P_\alpha) = \frac{1}{2}\,L^{AB}(P_\alpha)\,L_{AB}(P_\alpha) \ .
\end{align}
Our defect OPE block \eqref{Defect_OPE_Block} is easily seen to satisfy the equation as the integrand depending on the frame vectors itself is the solution to the quadratic Casimir equation
\begin{align}\label{Casimir_Eq_DO}
	\left(L^2(P_\alpha) + \CC_{\Delta, l}\right) \langle \CD^{(m)}( P_\alpha) \,\CO_{\Delta,l}(X,Z)\rangle = 0 \ .
\end{align}
This relation can be verified by using the expression \eqref{One_Point_Defect}, or the repeated use of the conformal invariance of the correlator,
\begin{align}\label{Defect_Shadow_Invariance}
\begin{aligned}
	\hat J_{AB}\, \langle \CD^{(m)}( P_\alpha) \,\CO_{\Delta,l}(X,Z)\rangle &=  \left(L_{AB}(P_\alpha) + J_{AB} (X) \right) \langle \CD^{(m)}( P_\alpha) \,\CO_{\Delta,l}(X,Z)\rangle \\
    &= 0 \ .
\end{aligned}
\end{align}
The conformal Casimir equation \eqref{Defect_OPE_Eigen} can be interpreted as a Klein-Gordon equation when the block is regarded as a scalar field on the moduli space as we will discuss in section \ref{ss:Moduli}.

In addition to the conformal Casimir equation \eqref{Defect_OPE_Eigen}
there are a set of constraint equations that follow from the invariance of the correlator \eqref{Defect_Shadow_Invariance} under the conformal group.
Let us define operators quadratic in the conformal generators by
\begin{align}\label{RangeOperator}
	\hat C_{ABCD} \equiv \hat J_{AB} \hat J_{CD} - \hat J_{AC} \hat J_{BD} + \hat J_{AD} \hat J_{BC} \ ,
\end{align}
for $1\le A < B < C < D \le d+2$.
When acted on a scalar function of the coordinate vector $X$ the spin part $\hat S_{AB}$ vanishes while the orbital part $\hat L_{AB}$ takes the differential form \eqref{L_X}, resulting in the operators $\hat C_{ABCD}$ being trivially zero.
Thus we obtain the following constraints for the scalar primary fields \cite{deBoer:2016pqk},
\begin{align}
	C_{ABCD}(X)\, \CO_{\Delta}(X) = 0 \ .
\end{align}
Acting on the defect OPE blocks for $l=0$ and applying the relation \eqref{Defect_Shadow_Invariance} twice, we obtain
\begin{align}\label{RangeEquation}
\begin{aligned}
	C_{ABCD}(P_\alpha) \, \CB^{(m)}[P_\alpha, \CO_{\Delta}] &= \frac{1}{\CN_\Delta}\, \int D^d X \,\tilde\CO_{d-\Delta} (X)\,C_{ABCD}(X)\,\langle \CD^{(m)}(P_{\alpha}) \,\CO_{\Delta}(X)\rangle \ ,\\
    &= \frac{1}{\CN_\Delta}\,\int D^d X \,\left(C_{ABCD}(X)\,\tilde\CO_{d-\Delta} (X)\right)\,\langle \CD^{(m)}(P_{\alpha}) \,\CO_{\Delta}(X)\rangle\ , \\
    &= 0 \ .
\end{aligned}
\end{align}
Note that the blocks with non-zero spins, $l\neq 0$, do not necessarily vanish when acted by the operator $C_{ABCD}$ as the spin part makes a non-vanishing contribution.

\paragraph{Dual description}
It is worth commenting on the dual representations of the defect OPE blocks \eqref{Defect_OPE_Block} with the dual defects.
They are simply build by replacing the defect $\CD^{(m)}(P_\alpha)$ with its dual $\CD^{(d+2-m)}(\tilde P_{\alpha})$ in \eqref{Defect_OPE_Block}:
\begin{align}\label{Defect_OPE_Block_dual}
	\CB^{(m)}[\tilde P_{\tilde\alpha}, \CO_{\Delta, l}] = \frac{1}{\CN_{\Delta, l}}\int D^d X \,\tilde\CO_{d-\Delta, l} (X, D_Z)\,\langle \CD^{(d+2-m)}(\tilde P_{\tilde\alpha}) \,\CO_{\Delta, l}(X, Z)\rangle \big|_{M=e^{2\pi i \Delta}} \ .
\end{align}
The dual representation is equivalent to the original one and is also a solution to the conformal Casimir equation \eqref{Defect_OPE_Eigen} and constraint equations \eqref{RangeEquation}, where the conformal generator should be written in the dual frames,
\begin{align}
	L_{AB}(\tilde P_{\tilde\alpha}) = \sum_{\tilde\alpha = 1}^{d+2 -m} \left( \tilde P_{A\tilde\alpha} \frac{\partial}{\partial \tilde P^B_{\tilde \alpha}} - \tilde P_{B\tilde\alpha} \frac{\partial}{\partial \tilde P^A_{\tilde\alpha}}\right) \ .
\end{align}
We will find it useful to move to the dual description when making a precise relation of our defect OPE block for a codimension-two defect to the OPE block defined by \cite{Czech:2016xec,deBoer:2016pqk} in the next subsection.

\subsection{Codimension-two defect}\label{ss:OPE_block}
In the case of $m=2$, the defect OPE block was originally introduced as the OPE block by \cite{Czech:2016xec,deBoer:2016pqk} in analyzing the causal structure of entanglement in CFT.
Their integral representation of the block includes the correlation function of the shadow operator and two virtual operators located on the tips of the causal diamond of a spherical entangling surface in Lorentzian signature.
In this subsection, we will reproduce their result from our defect OPE blocks by identifying their virtual operators with the dual defect of codimension-$d$ for the given codimension-two defect in the Euclidean signature.

As seen in section \ref{ss:Codim-two} the dual of a codimension-two defect is of codimension-$d$ and consists of a pair of local operators located at timelike separated points $X_1$ and $X_2$.\footnote{These are denoted with tildes, $\tilde X_1$ and $\tilde X_2$, in \eqref{Dual_Defect_m=2}.}
The defect OPE block in the dual frame \eqref{Defect_OPE_Block_dual} is the integral involving a three-point function in the integrand:
\begin{align}\label{Conf_Integral3}
	\CB^{(2)}[P_\alpha, \CO_{\Delta, l}] = \frac{1}{\CN_{\Delta,l}}\int D^d X_3\, \tilde\CO_{d-\Delta, l} (X_3, D_3)\,\langle \Phi (X_1)\, \Phi(X_2) \,\CO_{\Delta, l}(X_3, Z_3)\rangle \bigg|_{M=e^{2 \pi i \Delta}} \ .
\end{align}
Assuming the codimension-two defect has vanishing conformal dimension, the dual defect $\Phi(X)$ does so, and the three-point function \eqref{3-pt} becomes
\begin{align}
	\langle \Phi (X_1)\, \Phi(X_2)\,\CO_{\Delta, l}(X_3, Z_3) \rangle = a_3 \left(\frac{X_{12}}{X_{23}\, X_{31}}\right)^{(\Delta + l)/2} \left[ \frac{(Z_3\cdot X_1)X_{23} - (Z_3\cdot X_2)X_{13}}{X_{12}}\right]^l \ ,
\end{align}
with a normalization factor $a_3$.

To simplify the expression and make contact with the OPE block given by \cite{Czech:2016xec,deBoer:2016pqk}, we introduce a vector $K^A(X_3)$ as a function of $X_3$ in the embedding space,
\begin{align}
	K^A(X_3) \equiv \,\frac{X_1^A  X_{23} - X_2^A X_{13}}{X_{12}} \ ,
\end{align}
whose norm is given by
\begin{align}
	K^2 = \frac{X_{23}\, X_{31}}{X_{12}} \ .
\end{align} 
With this vector, we can recast the defect OPE block in a neat form:
\begin{align}\label{Conf_Integral4}
	\CB^{(2)}[P_\alpha, \CO_{\Delta, l}] = \frac{a_3}{\CN_{\Delta,l}} \int D^d X_3\, \tilde\CO_{d-\Delta, l} (X_3, D_3)\,|K|^{-\Delta - l} \,(Z_3 \cdot K)^l\, \bigg|_{M=e^{2 \pi i \Delta}}\ .
\end{align}

We further use the formula \eqref{Contracted_Tensors} to pull back the block from the embedding space to the coordinate space by substituting $F(X_3,D_3)=\tilde\CO_{d-\Delta, l}(X_3,D_3)$ and $G(X_3,Z_3) = (Z_3\cdot K)^l$.
The integral measure $D^d X_3$ in the embedding space is reduced to the canonical measure on flat space in the Poincar{\'e} section $X = (1,x^2, x^i)$, hence we have
\begin{align}\label{OPE_block}
	\CB^{(2)}[P_\alpha, \CO_{\Delta, l}] = 
    a_3'\int d^d \xi\, |k|^{-\Delta - l} k^{a_1}\cdots k^{a_l}\, (\tilde\CO_{d-\Delta})_{a_1\cdots a_l} (\xi) \,\bigg|_{M=e^{2 \pi i \Delta}} \ ,
\end{align}
where we redefined the integration variable to $\xi \equiv x_3$ and absorbed a numerical factor arising in the pull back into a constant $a_3'$.
The coordinate space vector $k^{a}$ in the integrand is pulled back from the embedding space vector $K^A$.
Indeed $k^a$ is the conformal Killing vector for the codimension-two defect.
One way to see this is to represent $k^a$ in the Poincar{\'e} section,
\begin{align*}
	k^a(\xi) = \frac{1}{(x_1 - x_2)^2} \left[ (x_1 - \xi)^a (x_2 - \xi)^2 - (x_2 - \xi)^a (x_1 - \xi)^2 \right] \ ,
\end{align*}
which fixes the positions of the two operators at $\xi = x_1$ and $\xi = x_2$ where $k^a$ vanishes.
It is an easy task to check whether it is a conformal Killing vector.

\paragraph{Comparison to the OPE block}

Our defect OPE block \eqref{OPE_block} resembles to the OPE block given by \cite{Czech:2016xec,deBoer:2016pqk} and they are actually related by the following replacements,
\begin{align}
	\CO_{\Delta} ~ \leftrightarrow ~ \tilde \CO_{d-\Delta} \ , \qquad \Delta ~  \leftrightarrow ~ d - \Delta \ .
\end{align}
In other word, the OPE block of \cite{Czech:2016xec,deBoer:2016pqk} is equivalent to our defect OPE block of a codimension-two operator in the dual description \eqref{Defect_OPE_Block_dual}.

For establishing the complete equivalence between ours and theirs, it remains to show the conformal generator $L_{AB}$ acted on the coordinates $X_1$ and $X_2$ takes the form,
\begin{align}
	L_{AB}(P_\alpha) = \sum_{\alpha = 1,2} \left( X^A_\alpha \frac{\partial}{\partial X^B_\alpha} - X^B_\alpha \frac{\partial}{\partial X^A_\alpha}\right) \ .
\end{align}
One can check it explicitly using the relation
\begin{align}
	X_1 = P_1 + i P_2 \ , \qquad X_2 = P_1 - i P_2 \ ,
\end{align}
between the frame vector specifying a codimension-two defect and the coordinates of the two dual local operators, which has been derived already in \eqref{FrameToDualCoord}.
The monodromy prescription on the frame vectors acts on $X_{1,2}$ as $X_{1,2} \to e^{2\pi i \Delta} X_{1,2}$, which is precisely the condition proposed by \cite{SimmonsDuffin:2012uy} for the conformal block of four local operators.

\subsection{Moduli space of conformal defects}\label{ss:Moduli}
A codimension-$m$ defect has a moduli space in CFT$_d$, denoted by $\CM^{(d,m)}$, parametrized by its size and the position of the center. 
The moduli space has a coset space structure,
\begin{align}\label{Defect_Moduli}
	\CM^{(d,m)} = \frac{SO(d+1,1)}{SO(m)\times SO(d+1-m,1)} \ ,
\end{align}
as the full conformal group $SO(d+1,1)$ acts on the defect while the subgroups $SO(m)$ and $SO(d+1-m,1)$ are the stabilizer group acting as the rotation around and the conformal transformation along the defect.
The dimension of the moduli space,
\begin{align}
	\text{dim}\, \CM^{(d,m)} = m(d+2 - m) \ ,
\end{align}
is invariant under the exchange $m \leftrightarrow d+2 - m$.
This invariance is also manifest in Lorentzian signature, and suggests a duality relation between defects of codimension-$m$ and $(d+2-m)$ as  detailed in section \ref{ss:duality}.

Our defect OPE block $\CB^{(m)}$ is a function on the moduli space $\CM^{(d,m)}$ and satisfies the conformal Casimir equation \eqref{Defect_OPE_Eigen}, which can be viewed as the Klein-Gordon equation on the moduli space in the following way.

The moduli space $\CM^{(d,m)}$ is an example of a symmetric coset space $G / H$ where $H$ is a subgroup of a Lie group $G$ whose Lie algebra $\mathfrak{g}$ is a direct sum of the Lie algebra $\mathfrak{h}$ of $H$ and a subspace $\mathfrak{m}$,
\begin{align}
\mathfrak{g} = \mathfrak{h}\, \oplus\,\mathfrak{m} \ ,
\end{align}
satisfying the relations,
\begin{align}
	 [\mathfrak{h}, \mathfrak{h}] \subset \mathfrak{h} \ , \qquad [\mathfrak{h}, \mathfrak{m}] \subset \mathfrak{m} \ , \qquad [\mathfrak{m}, \mathfrak{m}] \subset \mathfrak{h} \ .
\end{align}
A (pseudo-)Riemannian connection of the $G$-invariant metric on $G/H$ is descended from the invariant Cartan-Killing metric on $G$.
The $G$-invariant Laplacian $\square_{G/H}$ can be defined on $G / H$ with the (pseudo-)Riemannian connection, for which the following formula holds \cite{Pilch:1984xx,Camporesi:1990wm}:
\begin{align}\label{Coset_Laplacian}
\square_{G/H} f = \left[C_G - C_H \right]f \ .
\end{align}
Here $C_G$ and $C_H$ are the eigenvalues of the quadratic Casimir operators of the Lie group $G$ and $H$ for a harmonic function $f$ on $G / H$.
Hence, for a function invariant under $H$, the quadratic Casimir operator $\hat C_G$ on $G$ can be identified with the Laplacian $\square_{G/H}$ on the coset space.

The defect moduli space $\CM^{(d,m)}$ with the coset structure \eqref{Defect_Moduli} is indeed a symmetric coset space with $G=SO(d+1,1)$ and $H = SO(m) \times SO(d+1-m,1)$.
Applying the relation \eqref{Coset_Laplacian} to the present case, the Casimir element $L^2$ of the conformal group can be considered as the Laplacian $\square_{\CM^{(d,m)}}$ on the defect moduli space:
\begin{align}\label{Casimir_Laplacian}
-L^2 ~\longleftrightarrow ~\square_{\CM^{(d,m)}}\ .
\end{align}
The minus sign in the left hand side is due to the anti-hermicity of the generator in our convention.
This is the key relation to generalize the argument of \cite{Czech:2016xec} to higher-codimensional defects.
It allows an interpretation such that the equation \eqref{Defect_OPE_Eigen} is the Klein-Gordon equation for the defect OPE block $\CB^{(m)}$ when viewed as a scalar field on the moduli space $\CM^{(d,m)}$.

\section{Constructing local AdS operators from conformal defects}\label{ss:Radon}
In this section, we construct an AdS scalar field from the defect OPE blocks by exploiting the Radon transform between the AdS space and the moduli space $\CM^{(d,m)}$ of conformal defects.
We begin with commenting on the isomorphism between $\CM^{(d,m)}$ and  the moduli space of totally geodesic submanifolds in AdS, followed by
a brief review of the mathematical structure of the Radon transform in the group theoretical language \cite{helgason}.
We examine the constraint equations for a Radon transformed field, which allows a natural identification of the defect OPE block in the scalar representation with an AdS scalar field.
Finally we employ the inversion formula of the Radon transform to derive a formula of reconstructing an AdS scalar field from the defect OPE block, and discuss the equivalence to the Euclidean version of the HKLL formula \cite{Hamilton:2006fh} in the AdS/CFT correspondence.

\subsection{Conformal defects and totally geodesic submanifolds in hyperbolic space}\label{ss:Defect_Submanifold}
Associated to a codimension-$m$ defect in CFT$_d$ is a unique submanifold $\gamma^{(m)}$ of the same codimension that is anchored on the defect $\CD^{(m)}$ at the boundary of the AdS$_{d+1}$ space (see figure \ref{fig:Geodesics}).
This is most easily seen by recognizing that the conformal group $SO(d+1,1)$ is the isometry group $\text{Isom} (\text{AdS}_{d+1})$ of the (Euclidean) AdS$_{d+1}$ space and $SO(m)\times SO(d+1-m,1)$ is the stabilizer group $\text{Stab}( \gamma^{(m)}\in \text{AdS}_{d+1})$ of a totally geodesic submanifold $\gamma^{(m)}$ of codimension-$m$.
The moduli space of the submanifold $\gamma^{(m)}$ is therefore isomorphic to the defect moduli space:
\begin{align}\label{DModuli_AdS_Relation}
\CM^{(d,m)} = \frac{\text{Isom} (\text{AdS}_{d+1})}{\text{Stab}( \gamma^{(m)}\in \text{AdS}_{d+1})} \ .
\end{align}

Another way to see their relation is to show that the position of a submanifold $\gamma^{(m)}$ is fixed by the frame vector $P_\alpha$ of the corresponding conformal defect.
The embedding space formalism is suitable for this purpose as the AdS$_{d+1}$ space can be realized as a hypersurface satisfying the relation,
\begin{align}\label{AdS_Def}
	Y^2 = - \ell^2_\text{AdS} \ ,
\end{align}
where $Y$ is a vector in the embedding space and $\ell_\text{AdS}$ is the radius of the AdS space.
The bulk coordinate $Y$ can be given an explicit parametrization in several patches.
For example in the Poincar{\'e} patch, it takes the form,
\begin{align}\label{AdS_Poincare}
	Y = (Y^{+},Y^{-},Y^i)=\left( \frac{\ell_\text{AdS}^2}{z} , \frac{z^2 + x^2}{z}, \ell_\text{AdS} \frac{x_i}{z}  \right) \ ,
\end{align}
where $z$ is the holographic coordinates with the range $0\le z \le \infty$ and $x^i~(i=1,\cdots,d)$ are the flat space coordinates on the boundary $\BR^d$ at $z=0$.
A defect is characterized by the $m$ frame vectors $P_\alpha~(\alpha = 1, \cdots, m)$ as a codimension-$m$ hypersurface satisfying \eqref{Defect_Def}.
Similarly, we can specify a totally geodesic submanifold of codimension-$m$ in the AdS$_{d+1}$ by the conditions,
\begin{align}\label{Gamma_Def}
	 P_\alpha \cdot Y = 0 \ .
\end{align}
This realization manifests the fact that the stabilizer group $\text{Stab}( \gamma^{(m)}\in \text{AdS}_{d+1})$ of the submanifold is $SO(m)\times SO(d+1-m,1)$.
We can approach to the boundary of the AdS space by rescaling the bulk coordinate $Y$ as
\begin{align}\label{Bulk_to_Bdy}
	Y = \lambda X \ , \qquad \lambda \to \infty \ .
\end{align}
Then the AdS space \eqref{AdS_Def} approaches to the null cone $X^2=0$ and the conditions \eqref{Gamma_Def} reduce to the definition of a codimension-$m$ defect \eqref{Defect_Def}.
In the case of the Poincar{\'e} patch, we can choose $\lambda = \ell_\text{AdS}^2/z$ and take the $z\to 0$ limit to reach to the boundary point $X = (1, x^2/\ell_\text{AdS}^2, x_i/\ell_\text{AdS})$ in the Poincar{\'e} section of the null cone.
The submanifold $\gamma^{(m)}$ is a hyperbolic space whose boundary is a sphere as illustrated in figure \ref{fig:Geodesics}.
This construction guarantees the uniqueness of the submanifold $\gamma^{(m)}$ for a given conformal defect and explains why they have the same moduli space.

\definecolor{chameleon}{rgb}{0.30588, 0.60392, 0.023529}
\definecolor{lightblue}{rgb}{0.15,0.35,0.75}
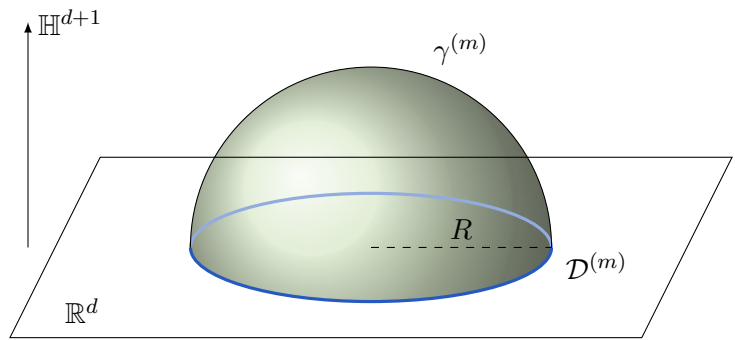
\begin{figure}[t]
\centering
\begin{tikzpicture}[scale=1.2]

    \shade[shading=radial, inner color=chameleon!80!white, outer color=chameleon!20!white, opacity=0.70] (2,0) arc (0:360:2cm and 0.6cm); 
    \tikzset{>=latex}
    
	\begin{scope}
 	    \clip(-3,3) -- (-3,0) -- (-2,0) arc (180:360:2cm and 0.6cm) -- (3,0) -- (3,3) -- (-3,3);
		\shade[ball color=chameleon!20!white, opacity=0.70] (0,0) circle (2cm);
	\end{scope}
	
    \draw[very thick, lightblue!50] (2,0) arc (0:180:2cm and 0.6cm);
    \draw[very thick, lightblue] (-2,0) arc (180:360:2cm and 0.6cm);

    \draw (2,0) arc (0:180:2cm and 2cm);

    \draw[dashed] (0,0)--(2,0);
	\node at (1,.2) {$R$};
    
    \draw (-3,1) -- (-4, -1) -- (3, -1) -- (4, 1) -- (-3,1);
    
    \draw[->] (-3.8, 0) -- (-3.8, 2.5) node[right] {$\BH^{d+1}$};
    \node at (1,2.2) {$\gamma^{(m)}$};
    \node at (2.5, -0.2) {$\CD^{(m)}$};
    \node at (-3.2, -0.7) {$\BR^d$};
\end{tikzpicture}
\caption{A totally geodesic submanifold in $\BH^{d+1}$ that is anchored on a conformal defect in $\BR^d$}
\label{fig:Geodesics}
\end{figure}

The isomorphism \eqref{DModuli_AdS_Relation} lets us  identify a totally geodesic manifold $\gamma^{(m)}$ with a conformal defect living on its boundary.
It will be used in conjugation with the Radon transform to translate the characteristics of the defect OPE blocks in CFT$_d$ to the physics in the AdS$_{d+1}$ space in the following subsections.

\subsection{Radon transform}
The AdS$_{d+1}$ spacetime is equal to the $(d+1)$-dimensional hyperbolic space $\mathbb{H}^{d+1}$ in the Euclidean signature.
We use these terms interchangeably from now on.

We first define the Radon transform on a coset space.
Let $G$ be a locally compact group, and $K, H$ be the subgroups of $G$.
We also define $L = K\cap H$ to be the intersection of them.  
A {\it double fibration} is the structure with two projections $p, \pi$, of the following form:
\begin{align}\label{Double_Fibration}
\xymatrix{
	&G/L \ar[ld]_p \ar[rd]^\pi &\\
X\equiv G/K& & \Xi \equiv G/H
	}
\end{align}
A map from $X$ to $\Xi$ is given by $\pi\circ p^{-1}$ such that $x \in X \mapsto \check{x}\subset \Xi$, and similarly from $\Xi$ to $X$ by $p\circ\pi^{-1}$ such that $\xi \in \Xi \mapsto \hat{\xi}\subset X$.
More concretely, associated to a point $x = g\,K$ on $X$ is the subspace given by
\begin{equation}\label{x_hat}
	\check{x} = \left\{g\,k\,H \, |\, k\in K\right\}\ ,
\end{equation}
and similarly for $\xi = g\, H$, $\hat{\xi} = \{ g\, h\, K 
\,|\, h\in H \}$.

Under these assumptions, we consider the Radon transform that is an integral transform from a function on $X$ ($\Xi$) to a function on $\Xi$ ($X$).\footnote{
The class of functions we will consider is analytic functions converging sufficiently rapidly to zero as the variable goes to infinity.
We have two reasons for this assumption.
One is to make the Radon (or its dual) transform well-defined, and
the other is to apply the result of \cite{ishikawa1997range} to the arguments in subsection \ref{ss:Intertwine}, where we will discuss the range characterization of the Radon transform.
For example, when $X=\text{AdS}_{d+1}$, we consider functions decreasing more rapidly than $(\cosh(r))^{-d/2}$ with $r$ the radial coordinate in the global section.
Since $(\cosh(r))^{-d/2}$ has the same convergence property as $(\sinh(r))^{-d/2}$ in $r\to\infty$, this class of functions are the normalizable modes in the AdS space.
The precise definition of this class of the analytic functions can be found in section 3 of \cite{ishikawa1997range}.
}
In order to define the integral on the coset spaces, we need measures on $K/L$ and $H/L$.
The existence of such measures is, in general, guaranteed when $L$ is compact and the transversality condition 
\begin{equation*}\label{transversality}
	K\cap H = K_H = H_K\,,\quad \text{with}\quad K_H \equiv\left\{k\in K \,\,|\,\,k\,H\cup k^{-1}\,H \subset HK \right\}\,,
\end{equation*} 
is satisfied.
In what follows, we consider the case where the space $X$ and $\Xi$ are the Riemannian manifold with the Riemannian measures.
We denote the measure on $K/L$ ($H/L$) by $dk_L$ ($dh_L$).
In this case, for arbitrary $\gamma \in G$, the Radon transform is defined by
\begin{equation}\label{Radon_def}
\hat{\phi}(\gamma\, H) = \int_{H/L}\,dh_L\,\phi(\gamma\, h\,K)\ , 
\end{equation}
for a continuous function $\phi$ on $X$.
The dual Radon transform is similarly defined by
\begin{align}\label{Dual_Radon_def}
	\check{f}(g\,K) = \int_{K/L} \,dk_L\, f(g\,k\,H)\ ,
\end{align}
for a continuous function $f$ on $\Xi$ and $g\in G$.

Note that the Radon transform of a function $\phi$ followed by the dual Radon transform does not necessarily return the original function, $\check{\hat \phi} \neq \phi$, while one can recover $\phi$ from $\hat \phi$ using an inversion formula in certain cases including ours.
We will give an explicit inversion formula and apply it to reconstruct a bulk scalar field propagating on the AdS$_{d+1}$ space from the defect OPE blocks in section \ref{ss:HKLL}.

We now turn to a specific case where $X$ is the Euclidean AdS$_{d+1}$ space and $\Xi$ the moduli space $\CM^{(d,m)}$ of codimension-$m$ conformal defects in CFT$_d$.
In this setting, the space $X$ is equivalent to a hyperbolic space $\BH^{d+1}$ whose coset representation is 
\begin{equation}
\mathbb{H}^{d+1} \simeq \frac{SO(d+1,1)}{SO(d+1)}\,.
\end{equation}
On the other hand, the moduli space $\mathcal{M}^{(d,m)}$ is the coset space defined by \eqref{Defect_Moduli}.
In setting the groups appearing in \eqref{Double_Fibration} to be
\begin{align}
G = SO(d+1,1)\ , \qquad K= SO(d+1)\ , \qquad H = SO(m)\times SO(d+1-m,1) \ ,
\end{align}
the AdS$_{d+1}$ space and the moduli space $\CM^{(d,m)}$ form a double fibration with $L = SO(m)\times SO(d+1-m)$.
In the matrix realization, each subgroup is embedded into $G= SO( d+1,1)$ as follows,
\begin{equation*}
	\begin{tikzpicture}
	\matrix(m)[matrix of math nodes,nodes in empty cells,
	ampersand replacement=\&,
	left delimiter={(},right delimiter={)},
	inner sep=0.8ex]
	{
		a_{11}\&\&\&\&\&\&\&\&\&\&\\
		\&a_{22}\&\&\&\&\&\&\&\&\&\\
		\&\& \&\&\&\&\&\&\&\&\\
		\&\&\&\&\&\&\&\&\&\&\\\
		\&\&\&\& \&\&\&\&\&\&\\
		\&\&\&\&\&\&\&\&\&\&\\
		\&\&\&\&\&\& \&\&\&\&\\
		\&\&\&\&\&\&\&\&\&\&\\
		\&\&\&\&\&\&\&\& \&\&\\
		\&\&\&\&\&\&\&\&\& a_{d+1,d+1} \&\\
		\&\&\&\&\&\&\&\&\&\& a_{d+2, d+2}\\
	};
	\node[left] at (-4, 1.5) {\textcolor{red!60}{$SO(m) \ni$}};
	\node[right] at (3.7, -1.2) {\textcolor{red!60}{$\in SO(d+1-m,1)$}};
	\node[left] at (-4, 0) {\textcolor{blue!60}{$SO(d+1)\ni$}};
	\node[right] at (4.5, 0) {{\Large $\in SO(d+1,1)$\ .}};
	\draw[dotted] (0,-0.5)--(-1.6, 0.7);
	\draw[dashed] (-3.7,-1.4)--(3.7,-1.4);

	\draw[dashed] (1.75, 2.2)--(1.75, -2.2);
	\draw[red!50, line width = 2pt, rounded corners] (-3.5, 2)--(-3.5, 0)--(-0.5, 0)--(-0.5, 2)--cycle;
	\draw[blue!50, line width = 2pt, rounded corners] (-3.6, 2.1)--(1.7, 2.1)--(1.7, -1.3)--(-3.6, -1.3)--cycle;
	\draw[red!50, line width =2pt, rounded corners] (-0.45,-0.1)--(-0.45, -2.2)--(3.5,-2.2)--(3.5, -0.1)--cycle;
	\end{tikzpicture}
\end{equation*}

The Radon transform defined by \eqref{Radon_def} in the present case is written as
an integration over a $(d+1-m)$-dimensional hyperbolic space as we have a coset space isomorphism,
\begin{align}
 H /L & = \frac{SO(m)\times SO(d+1-m,1)}{ SO(m)\times SO(d+1-m)} \ ,\nonumber\\
      & \simeq \frac{SO(d+1-m,1)}{SO(d+1-m)} \simeq \mathbb{H}^{d+1-m}. 
\end{align}
Note that the integration range $H/L = \mathbb{H}^{d+1-m}$ of the Radon transform from $X=\BH^{d+1}$ to $\Xi=\CM^{(d,m)}$ is a submanifold of the Euclidean AdS space $\mathbb{H}^{d+1}$.
Hence we are able to construct a scalar function $\hat \phi(\xi)$ on the moduli space by smearing a scalar field $\phi(x)$ on the AdS space over a submanifold $\hat{\xi}$ isomorphic to $\BH^{d+1-m}$,
\begin{align}
\hat{\phi}(\xi) = \int_{x\in \hat{\xi}} \,d^{d+1-m}h\, \sqrt{h}\,\phi(x) \ ,
\end{align}
where the volume element $\sqrt{h}\,d^{d+1-m}h$ is induced from $\BH^{d+1}$ onto $\mathbb{H}^{d+1-m}$.
Note that $\hat{\xi}$ is a totally geodesic submanifold corresponding to the point $\xi$ in $\Mdm$.
Performing the integrals for a function $\phi(x)$ in the Euclidean AdS$_{d+1}$ space over all such submanifolds results in a scalar function $\hat \phi(\xi)$ on the moduli space.

Similarly we can build a function $\check{f}(x)$ on the AdS space by smearing a function $f(\xi)$ on the moduli space $\Mdm$,
\begin{align}
	\check{f}(x) = \int_{\xi \in \check{x}} d\mu(\xi) \,f(\xi)\ ,
\end{align}
over a submanifold $\check x$ isomorphic to an oriented Grassmannian,
\begin{align}
	K/L = \frac{SO(d+1)}{SO(m) \times SO(d+1 - m)} \ ,
\end{align}
with the Riemannian measure $d\mu(\xi)$.
This formula provides us a practical method for reconstructing a scalar field on the AdS$_{d+1}$ space from a set of data of conformal defects in CFT$_d$.
This construction was undertaken by \cite{Czech:2016xec} to realize a scalar field propagating on the AdS$_{d+1}$ space from the OPE block of a scalar primary field in CFT$_d$.
We will extend this idea to the defect OPE block of a scalar primary and construct an AdS scalar field upon the identification $\hat\phi = \CB^{(m)}[\CO_\Delta]$ for any codimension-$m$ below.

\subsection{Intertwining property and constraint equations}\label{ss:Intertwine}

One of the salient features of the Radon transform is the intertwining property that gives a natural relation between a certain type of differential operators on $X$ and $\Xi$.
As we will see shortly, this property motivates us to identify a scalar field on the moduli space $\Mdm$ with the defect OPE block.

To begin with, we observe that the Radon transform commutes with the Lie group action,
\begin{equation}
 (\hat\phi)^{\tau(g)}  = \widehat{(\phi^{\tau (g)})} \qquad \text{for} \qquad g \in G\ ,
\end{equation}
where the $G$ action on a function $\phi$ is defined by
\begin{equation}
 \phi^{\tau(g)}(x) \equiv \phi(g^{-1}x) \ .
\end{equation}
Combined with the homomorphism from the Lie algebra to the differential operator,
\begin{equation}
Y\cdot \phi(x) \equiv \frac{d}{dt} \phi\left( e^{-tY}\cdot x\right)\Big|_{t=0}\ ,
\end{equation}
we verify the Radon transform intertwines the differential operators between $X$ and $\Xi$.

We now turn to apply the intertwining property in translating the constraint equations on $X$ to those on $\Xi$.
First, let us begin with the equation of motion of a scalar field on the AdS background,
\begin{align}
(\Box_\text{AdS} - m^2)\,\phi = 0 \ .
\end{align}
Since the Laplacian on the AdS space is the Laplace-Beltrami operator on the coset space $X = SO(d+1,1)/SO(d+1)$, the equation of motion is mapped by the Radon transform to the constraint equation on the moduli space $\Xi = \Mdm$, 
\begin{align}\label{EOM_on_Mdm}
(\Box_{\Mdm} - m^2)\,\hat{\phi} = 0\ .
\end{align}
As seen from the relation \eqref{Casimir_Laplacian}, the Laplacian on $\Mdm$ is equal to (minus) the quadratic Casimir operator,
thus the equation \eqref{EOM_on_Mdm} turns out to be the quadratic Casimir equation \eqref{Defect_OPE_Eigen} for the defect OPE block if we identify the Radon transformed field with the block,
\begin{align}\label{Radon_DOPE}
	\hat\phi = \CB^{(m)}[ \CO_\Delta ] \ ,
\end{align}
and the mass squared with the Casimir eigenvalue $\CC_{\Delta,0}$,
\begin{align}\label{mass_AdS}
	m^2 = \Delta(\Delta - d) \ .
\end{align}
This is the well-known relation between the mass of the AdS scalar field and the conformal dimension of the corresponding primary operator in the AdS/CFT correspondence.

When a function $\phi$ is annihilated by a differential operator generated by $g \in \mathfrak{g}$, the Radon transform $\hat{\phi}$ is also annihilated by the action of the operator corresponding to the same $g$.
Such an annihilation operator is important to characterize the range of the Radon transform.

The Radon transform intertwines functions on manifolds with different dimensions.
In our case, the moduli space $\CM^{(d,m)}$ is $m(d+2-m)$-dimensional, which is equal or greater than the dimension of the AdS$_{d+1}$ space,
\begin{align}
	\text{dim}\,\Mdm - \text{dim}\, \text{AdS}_{d+1} = (m-1)(d+1-m)\ ,
\end{align}
for $1\le m \le d$.
Hence a function on $\Mdm$ obtained by the Radon transform from the AdS$_{d+1}$ has to obey $(m-1)(d+1-m)$ constraint equations.
Such constraints are given by differential equations and are known as the range characterization of the totally geodesic Radon transform on the hyperbolic space (Corollary 11.4 in \cite{ishikawa1997range}).
Defining the range characterization operators by \eqref{RangeOperator}
a Radon transformed function $\hat \phi$ on the moduli space $\Xi = \Mdm$ satisfies the set of equations,\footnote{Another characterization for the Radon transform is given by a fourth-order differential equation
\begin{align}
	C^2\, \hat\phi = 0 \ , \qquad C^2 \equiv C_{ABCD}\,C^{ABCD} \ ,
\end{align}
whose solution has the same range as \eqref{RangeCharacterization} \cite{ishikawa1997range}.
}
\begin{align}\label{RangeCharacterization}
	C_{ABCD} \,\hat\phi = 0 \ , \qquad (1\le A < B < C < D \le d+2) \ .
\end{align}
These are $d(d^2-1)(d+2)/24$ second order differential equations that appear to overconstrain the range of a function on $\Mdm$ as only $(m-1)(d+1-m)$ equations out of them are expected to be independent \cite{deBoer:2016pqk}.
We are not aware of a systematic method to reduce the range characterization equations \eqref{RangeCharacterization} to a minimal set of equations.

The identification \eqref{Radon_DOPE} is consistent with the fact that the range characterization condition \eqref{RangeCharacterization} for $\hat\phi$ takes exactly the same form as the constraint equation \eqref{RangeEquation} for $\CB^{(m)}[P_\alpha, \CO_\Delta ]$.
This is the key observation made by \cite{Czech:2016xec} that allows us to reconstruct a bulk scalar field by smearing a boundary primary scalar operator, which we will turn into next in section \ref{ss:HKLL}.

\subsection{Inversion formula and HKLL construction}\label{ss:HKLL}

The Radon transform $\phi \to \hat \phi$ and its dual $f \to \check f$ are not inverse to each other, and we need an inversion formula to recover $\phi$ from $\hat \phi$.
We will see that an explicit inversion formula is available in our setup.
Applying the formula to a function $\phi$ on the AdS$_{d+1}$ space together with the identification \eqref{Radon_DOPE} between $\hat\phi$ and the defect OPE block $\CB^{(m)}$, we will show a bulk scalar field can be reconstructed from a primary scalar in CFT, which turns out to be the HKLL formula proposed by \cite{Hamilton:2006fh} in the AdS/CFT correspondence. 

Before stating our main claim, we revisit the double fibration structure of the Radon transform from a slightly different viewpoint to make the geometric meaning clearer.
We view $X =G/K \simeq \mathbb{H}^{d+1}$ as a Riemannian manifold, and pick up a point $o$ (or $K$ in $G/K$) as the origin of $X$.
Let $\Xi$ be the space of $k$-dimensional totally geodesic submanifolds in $\BH^{d+1}$, which can be represented as a coset $\Xi_p = G/H_p$ with $H_p ( = H )$ being the isotropy group fixing an element $\xi_p \in \Xi_p$ at distance $d(o, \xi_p) = p$ from the origin $o$.
The subscript $p$ indicates the geodesics are at distance $p$ from the origin.
Note that, in this notation, we naturally identify $\hat{\xi}_p$ with $\xi_p$ through the geometric realization.
$\check x$ can also be seen as the set of all geodesics at distance $p$ form $x$.
Adapting the double fibration construction to this coset gives the same definition of the Radon and dual Radon transforms as in the previous subsection.
\begin{align}
\xymatrix{
	&G/(L=K\cap H_p) \ar[ld] \ar[rd] &\\
X\equiv G/K& & \Xi_p \equiv G/H_p
	}
\end{align} 
In this setup, we can rewrite the Radon and dual Radon transforms as
\begin{equation}
	\hat{\phi}(\xi_p) = \int_{x \in \xi_p}\,dm(x)\, \phi(x)\ ,\qquad\check{f}(x) = \int_{\check{x}} \,d\mu (\xi)\,f (\xi)\ ,
\end{equation}
where we used the measures induced by the Riemannian metrics.
The original definitions \eqref{Radon_def} can be recovered by using the relation \eqref{x_hat} and the identification between $\xi_p$ and $\hat\xi_p$.

The inversion formula of the Radon transform reconstructs a function $\phi(x)$ on $\BH^{d+1}$ from the Radon transformed field $\hat \phi (\xi_p)$ \cite{helgason1990totally,helgason},
\begin{align}\label{inversion_formula}
	\phi (x) = -c_{k} \left[ \frac{d^k}{d (r^2)^k}\, \int_r^{\infty} dt\, (t^2 - r^2)^{k/2 -1}t^k (M^{s(t)} \hat\phi) (x) \right]_{r=1} \ ,
\end{align}
where $c_k = 2/(\pi^{k/2} \Gamma(k/2))$, $s(t) = \text{arccosh} (t)$. $M^{s(t)}$ is the {\it mean-value operator} defined by
\begin{equation}\label{mean-value}
	\left(M^p \hat{\phi}\right)(x = g\cdot o) \equiv \int_{K} \hat{\phi}(gkg^{-1}\cdot g\,\xi_p)\,dk\ ,
\end{equation}
where we pick a ``reference" geodesic $\xi_p$, while the right hand side is independent of the choice of $\xi_p$ after the integration over $K$ as the $K$ action on the reference geodesic generates the other geodesics at distance $p$ from the origin.
We relegate the proof of the inversion formula to appendix \ref{ap:Inversion}.

Having stated the inversion formula \eqref{inversion_formula} for the Radon transform, we are now in position to apply it to reconstruct an AdS scalar field from the defect OPE block under the identification \eqref{Radon_DOPE}.

\begin{figure}
\begin{center}
\tdplotsetmaincoords{0}{0}
\begin{tikzpicture}[tdplot_main_coords, scale = 0.9]
\shadedraw[opacity = .1] (0,0,0) circle[radius = 3];
\draw[] (0,0,4) circle[radius = 3];
\draw[] (0,0,-4) circle[radius = 3];
\draw[dashed] (3,0,4)--(3,0,-4); 
\draw[dashed] (-3,0,4)--(-3,0,-4); 
\draw[dashed, thick] (0,0,0) circle [radius = 1.23];
\draw[<->, thick] (0,0,0)--(1.23,0,0);
\node[below] at (0.6,0,0){\large $p$};
\begin{scope}[rotate around z = -45]
	\draw[thick, cyan!70](0,3,0) arc[start angle=180, end angle = 270, radius = 3];
	\node[below right] at (3,0,0) {\large $\xi_p$};
\end{scope}
\begin{scope}[rotate around z = -43]
   \draw[very thick, ->] (2.3,0,0) arc[start angle=0,end angle= 70, radius =2.5];
\end{scope}
\begin{scope}[rotate around z =30]
	\draw[thick, red!70](0,3,0) arc[start angle=180, end angle = 270, radius = 3];
    \node[above right] at (2.9,0,0) {\large $k\cdot \xi_p$};
\end{scope}
\node[right] at (3,0,0) {\large $k\in SO(d+1)$};
\node[left] at (-3,0,4) {\large $\mathbb{H}^{d+1}$};
\end{tikzpicture}
\caption{$k\in SO(d+1)$ action on $\xi_p$ in the mean-value operator $M^p$}
\label{figure:k_action}
\end{center}
\end{figure}
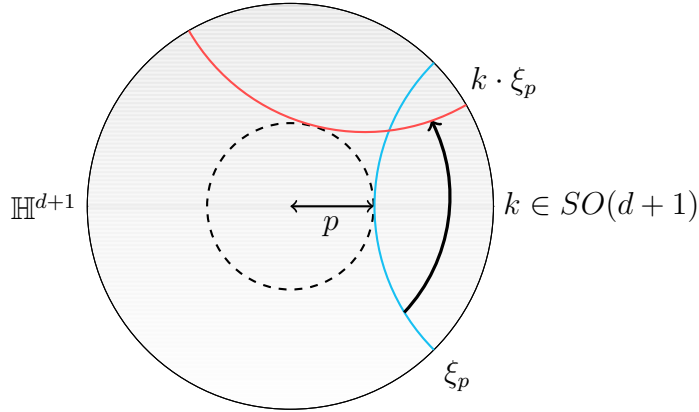

To simplify discussion, we start with the case when $g$ is the identity, $g= id$, in the formula \eqref{mean-value}.
Namely we consider the bulk operator $\phi$ at the center $x=o$.
See figure \ref{figure:k_action} for the illustration.
In this case, $M^{s(t)}$ averages the function $\hat{\phi}\left(\xi_{s(t)}\right)$ over a family of geodesics whose nearest distances from the origin are $s(t)$.

The distance $s(t)$ from the origin to a geodesics is determined only by the radius $R$ of the defect on its boundary through the relation,
\begin{equation}\label{rel_t_R}
	t = \frac{1}{\sin(R)}\ .
\end{equation}
This is easily obtained in the global section, but it does not depend on the choice of sections as the radius $R$ defined by \eqref{Defect_Radius} is written in the embedding space formalism.
Putting all together, we obtain the following identity,
\begin{align}
	\begin{aligned}
		\phi(o) &=  -c_{d+1-m} \left[ \frac{d^{d+1-m}}{d (r^2)^{d+1-m}}\, \int_r^{\infty} dt\, (t^2 - r^2)^{(d-1-m)/2} t^{d+1-m}\int_{SO(d+1)} dk\ \CB^{(m)}[k\cdot P_\alpha, \CO_{\Delta}] \right]_{r=1}\ ,\\
		&=- \frac{a_\Delta\,c_{d+1-m}}{\CN_\Delta} \int D^d X \,  \Biggl[ \frac{d^{d+1-m}}{d (r^2)^{d+1-m}}\, \int_r^{\infty} dt\, (t^2 - r^2)^{(d-1-m)/2}t^{d+1-m}  \\
		&\hspace{5cm}\cdot \int_{SO(d+1)} dk\frac{1}{\left[ ((k\cdot P)^\alpha \cdot X)((k\cdot P)_\alpha \cdot X)\right]^{\Delta/2}}\Biggr]_{r=1} \tilde\CO_{d-\Delta} (X)\,.
	\end{aligned}
\end{align}
Here, the action of $k\in K=SO(d+1)$ on $P^\alpha$ is induced by the natural restriction of the fundamental representation of $SO(d+1,1)$ to its subgroup $K$.\footnote{This action does not change the distance $t$ and the radius $R$, thus the reference point $\Omega$ must be in the fundamental representation of $SO(d+1,1)$.}

We act an element $g\in G=SO(d+1,1)$ on the origin $o$ to derive the following formula for constructing an AdS local field $\phi$ at a general position,
\begin{equation}\label{Bulk_reconstruction}
	\phi(Y) = -c_{d+1-m} \left[ \frac{d^{d+1-m}}{d (r^2)^{d+1-m}}\, \int_r^{\infty} dt\, (t^2 - r^2)^{(d-1-m)/2}t^{d+1-m} \int_{SO(d+1)} dk\ \CB^{(m)}[g\,k\cdot P_\alpha, \CO_{\Delta}] \right]_{r=1}\ .
\end{equation}
The embedding space vector $Y$ corresponds to a point $x = g\cdot o$ of the AdS$_{d+1}$ space.
The action of $g \in G $ on $P^\alpha$ is also given by the fundamental representation.
This formula \eqref{Bulk_reconstruction}, which may look abstract at first, consists of a few simple steps as follows:
\begin{itemize}
\item[1.] Fix a bulk point $Y = g\cdot o$ where we want to construct an AdS scalar field.
\item[2.] Pick one geodesic $\xi$ out of a family of $(d+1-m)$-dimensional totally geodesic submanifolds whose nearest distances from $Y$ are $\text{arccosh}(1/\sin R)$.
Then a spherical defect of radius $R$ anchors the geodesic $\xi$ on the boundary of $\BH^{d+1}$. 
\item[3.] Find the frame vectors $P_\alpha$ corresponding to the submanifold $\xi$.
\item[4.] Substitute the frame vectors $P_\alpha$ into the formula \eqref{Bulk_reconstruction}, and evaluate the integration.
\end{itemize}
This prescription is what we announced in Introduction for reconstructing an AdS scalar field from the defect OPE blocks of a scalar primary field in CFT.

We can rewrite the formula \eqref{Bulk_reconstruction} in a more insightful way in combination with the integral representation of the defect OPE block \eqref{Defect_OPE_Block}.
One can exchange the order of integrations and pull out the spatial integration to recast \eqref{Bulk_reconstruction} into the form,
\begin{equation}\label{HKLL_Shadow}
	\phi(Y) =\int D^d X \,  \tilde K_{d-\Delta}(Y|X)\,  \tilde\CO_{d-\Delta} (X)\, ,
\end{equation}
where $\tilde K_{d-\Delta}(Y|X)$ is the ``shadow" kernel defined by
\begin{align}
\begin{aligned}
    \tilde K_{d-\Delta}(Y|X) \equiv -\frac{a_\Delta\,c_{d+1-m}}{\CN_\Delta}\bigg[ &\frac{d^{d+1-m}}{d (r^2)^{d+1-m}}\, \int_r^{\infty} dt\, (t^2 - r^2)^{(d-1-m)/2}t^{d+1-m}\\
    & \cdot\int_{SO(d+1)} dk\,\frac{1}{\left[ ((gk\cdot P)^\alpha \cdot X)((gk\cdot P)_\alpha \cdot X)\right]^{\Delta /2}}\bigg]_{r=1}
    \,.
\end{aligned}
\end{align}
Carrying out this calculation on the right hand side appears to be difficult, but we can resort to the symmetry consideration to fix the form.
It is a scalar function of the two embedding vectors $X$ and $Y$, thus it can only depend on their inner product $X\cdot Y$.
The kernel must have a weight $-\Delta$ in $X$ to make the right hand side of \eqref{HKLL_Shadow} be a conformal integral.
Therefore the kernel results in the following unique form up to a factor,\footnote{In Lorentzian signature, one has to multiply $\Theta(-X\cdot Y)$ to the integrand of the kernel to respect the causality in the bulk space \cite{Xiao:2014uea}, which is absent after the Wick rotation.}
\begin{equation}\label{HKLL_kernel_shadow}
	\tilde K_{d-\Delta}(Y|X) = (-2 X\cdot Y)^{-\Delta}\ . 
\end{equation} 
This is equivalent to the Euclidean version of the HKLL formula \cite{Hamilton:2006fh} if we replace $\Delta$ and $\tilde\CO_{d-\Delta}$ with $d-\Delta$ and $\CO_\Delta$, respectively.
Indeed we can switch them using the dual representation \eqref{Defect_OPE_Block_flipped} of the defect OPE block $\CB^{(m)} [P_\alpha, \CO_{\Delta}]$, and arrive at the familiar form,
\begin{equation}\label{HKLL}
	\phi(Y) =\int D^d X \,  K_{\Delta}(Y|X) \, \CO_{\Delta} (X)\, ,
\end{equation}
with the kernel,
\begin{equation}\label{HKLL_kernel}
 K_{\Delta}(Y|X) = (-2 X\cdot Y)^{\Delta-d}\ . 
\end{equation} 
Equivalently one can go more directly from one to the other expression by performing the shadow transformation,
\begin{align}
	\begin{split}
		\int D^d X \,  \frac{1}{(-2 X\cdot Y)^{\Delta}} \tilde\CO_{d-\Delta} (X) &\propto \int D^d Z\,\int D^d X \,  \frac{1}{(-2 X\cdot Y)^{\Delta}}  \frac{1}{(-2X\cdot Z)^{d-\Delta }}\,\CO_{\Delta}(Z)\ ,\\
			&\propto \int D^d Z\, \frac{1}{(-2Z\cdot Y)^{d-\Delta}}\,\CO_{\Delta}(Z) \ ,
	\end{split}
\end{align} 
where we used the formula for the conformal integral \cite{SimmonsDuffin:2012uy},
\begin{equation}
\int D^d X \,  \frac{1}{(-2 X\cdot Y)^{\Delta}} \frac{1}{(-2X\cdot Z)^{d-\Delta }} = \frac{\pi^{d/2}\Gamma(\Delta-d/2)}{\Gamma(\Delta)}\frac{(-Y^2)^{d/2-\Delta}}{(-2Z\cdot Y)^{d-\Delta}}\,.
\end{equation}
with the relation for the AdS vector, $Y^2 = -\ell_\text{AdS}^2$.

The above argument shows the equivalence of \eqref{HKLL_Shadow} and \eqref{HKLL} arises from the two equivalent descriptions of the defect OPE block discussed in section \eqref{ss:IntegralRep}, where
we imposed the monodromy condition to remove the shadow block with an undesirable behavior in the small radius limit.
Here we also want to pick up in the reconstruction formula \eqref{Bulk_reconstruction} an appropriate contribution with the expected asymptotics near the boundary in the Poincar{\'e} section,
\begin{align}\label{Boundary_Behavior}
		\phi (Y) \to z^\Delta\,\CO(X) \ ,  \qquad z \rightarrow 0\ .
\end{align}
As prescribed in section \ref{ss:Defect_Submanifold}, the bulk point $Y$ approaches to the boundary point $X$ by rescaling $Y = \lambda\, X$ in the $\lambda \to \infty$ limit, where the expression \eqref{HKLL_Shadow} reproduces the asymptotics \eqref{Boundary_Behavior},
\begin{align}
 \phi(Y) & = \lambda^{-\Delta} \int D^d X_0 \frac{1}{(-2X_0 \cdot X)^{\Delta}}  \tilde\CO_{d-\Delta} (X_0)  \ ,\nonumber \\
 & = \lambda^{-\Delta}\,\CO_{\Delta}(X) \ , \nonumber \\
 & \xrightarrow[\lambda \to \infty]{} z^{\Delta}\,\CO_\Delta (X) \ .
\end{align}
Here $X$ approaches to a null vector in the last line and we use the identification between the scaling parameter $\lambda$ and the inverse of the AdS coordinate $1/z$ made below \eqref{Bulk_to_Bdy}.
On the other hand, repeating the same argument shows the formula \eqref{HKLL} has the different asymptotics,
\begin{align}
\phi(Y) \sim z^{d-\Delta}\,\tilde\CO_{d-\Delta}(X)\ .
\end{align}
The difference between the two asymptotic behaviors can be distinguished by the monodromy $M:Y \rightarrow e^{-2 \pi i}Y$ just like the defect OPE block.
Reassuringly, the monodromy condition we imposed on the defect OPE block in \eqref{Defect_OPE_Block} guarantees the correct asymptotics \eqref{Boundary_Behavior} of the AdS scalar field in the light of the AdS/CFT correspondence.

The scalar function $\phi(Y)$ satisfies the Klein-Gordon equation because of the kernel $K_\Delta (Y | X)$ (and the shadow kernel $\tilde K_{d-\Delta}(Y|X)$) being the propagator of the Klein-Gordon field in the AdS space, which can be seen from the explicit form \eqref{HKLL_kernel} (and its shadow \eqref{HKLL_kernel_shadow}).
Instead, we can argue that $\phi(Y)$ is a scalar field propagating on the AdS space by reminding the fact that the Radon transform intertwines the quadratic Casimir equation for the defect OPE block with the Klein-Gordan equation as we have discussed in section \ref{ss:Intertwine}.

Before concluding this section, we want to contrast our reconstruction formula \eqref{Bulk_reconstruction} with the one presented in \cite{Czech:2016xec}.
We used the inversion formula of the Radon transform from $\Mdm$ to $\mathbb{H}^{d+1}$ in order to make manifest the covariance of the reconstructed scalar field in the AdS$_{d+1}$ space.
If we would just need the value of the scalar field on a specified point, we could have chosen a time slice $\mathbb{H}^d$ including the point, and adopted the inversion formula on the slice for the reconstruction.
In other words,
we could have assumed defects were constrained on a constant Lorentzian time slice by fixing one of the frame vector $P_1$ to be parallel to the time direction, and applied the inversion formula from $\CM^{(d-1,m-1)}$ to $\mathbb{H}^d$ to derive another formula,
\begin{equation}
\phi(o) =  -c_{d+1-m} \left[ \frac{d^{d+1-m}}{d (r^2)^{d+1-m}}\, \int_r^{\infty} dt\, (t^2 - r^2)^{(d-1-m)/2} t^{d+1-m}\int_{SO(d)} dk\ \CB^{(m-1)}[k\cdot P_\alpha, \CO_{\Delta}] \right]_{r=1}\ .
\end{equation}
Regarding the block $\CB^{(m-1)}$ as an element in $\Mdm$ by the restriction $\CB^{(m-1)} \propto \CB^{(m)}|_{P_1\text{:fix}}$, we might allow ourselves to think of $\phi$ as a scalar function not on the time slice $\mathbb{H}^d$ but on the whole space $\mathbb{H}^{d+1}$.
This type of reconstruction was employed in \cite{Czech:2016xec} and exemplified for $d=m=2$, which is reproduced in appendix \ref{ap:HKLL} for comparison to our results.
The covariance of the reconstructed scalar field on the AdS space in their approach appears to be not a consequence but rather an assumption, while it is manifest in our approach.

\section{Discussion}\label{ss:Discussion}
In this paper, we initiated the detailed studies of the OPE structure for conformal defects with a view toward a better understanding of their universal aspects characterized by the conformal symmetry and enhancing the holographic dictionary for non-local objects in the AdS/CFT correspondence.
In this section we list open problems and possible applications of the defect OPE blocks, which we hope to address in future work.

\paragraph{Analytic continuation to Lorentzian signature}
Most of our analyses were carried out in Euclidean CFTs to take full advantage of the embedding space and the shadow formalism.
A price to pay is the contamination of the shadow contribution in the defect OPE block and the ambiguity of the analytic continuation to Lorentzian signature.
To overcome the former issue, we imposed the monodromy condition on the defect OPE block in Euclidean signature, whose implication was not clear at all after the Wick rotation.
Comparing our prescription for the codimension-two case with the OPE block \cite{Czech:2016xec,deBoer:2016pqk},
we speculate imposing the monodromy condition corresponds to restricting the integration range of the block to the interior of the causal domain $D(\CD^{(m)})$ for the defect operator in Lorentzian signature.

More precisely, we propose that the Lorentzian version of the defect OPE block is given by the Wick rotation of the Euclidean block \eqref{Defect_OPE_Block} with the integration range restricted to $D(\CD^{(m)})$,
\begin{align}\label{Lorentzian_DOPEB}
\CB^{(m)}[P_\alpha, \CO_{\Delta,l}]_\text{Lorentzian} = \frac{1}{\CN_{\Delta, l }} \int_{D(\CD^{(m)})} D^d X \,\tilde\CO_{d-\Delta,l} (X,D_Z)\,\langle \CD^{(m)}( P_\alpha) \,\CO_{\Delta,l}(X,Z)\rangle \ .
\end{align}

The restriction of the integration range amounts to guarantee the causality of the HKLL formula in the Lorentzian AdS space when $m=2$. 
While the causal domain of a codimension-two defect is well understood as a causal diamond, the meaning of the causal domain of a higher-codimensional defect is less clear.
We leave the question open whether the Lorentzian defect OPE block \eqref{Lorentzian_DOPEB} will lead to the correct HKLL formula maintaining the causality in the AdS space.

\paragraph{Radon transform of higher spin fields}
Formally nothing prevents us from defining the defect OPE block of a spinning defect.
It would satisfy a quadratic Casimir equation whose Radon transform may be interpreted as an equation of motion for a higher spin field.
The Radon transform of a bundle valued field is defined in \cite{branson1994bundle}, but only for a codimension-one totally geodesic hyperplane in AdS.
We are not aware of any generalization of the Radon transform to a general case except the work \cite{Czech:2016tqr} concerned with the spin two case.

While we have nothing concrete to say about the Radon transform for higher spin fields, we note the following observation that motivates us to consider spinning defects as the counterparts of higher spin fields if such a Radon transform exists.
In section \ref{ss:Moduli}, we associated the Laplacian on the coset space $G/H$ to the quadratic Casimir operators of the groups $G$ and $H$ via the identity \eqref{Coset_Laplacian}.
A similar formula is known for the Lichnerowicz operator $\triangle_{G/H}$ applied to a tensor harmonic $f$ on a (compact) coset space \cite{Pilch:1984xx,boucetta2009lichnerowicz}:
\begin{align}\label{Lichnerowicz}
	\triangle_{G/H} f = C_G\,f \ .
\end{align}
Note that the Lichnerowicz operator differs from the Laplacian when acted on a symmetric traceless tensor $f_{\mu_1\cdots \mu_l}$,
\begin{align}
\begin{aligned}
	\triangle_{G/H}f_{\mu_1\cdots \mu_l} &= \square_{G/H} f_{\mu_1\cdots \mu_l} + \sum_{i=1}^l R^\alpha_{~\mu_i}\,f_{\mu_1\cdots \alpha\cdots \mu_l} 
        + \sum_{i,j=1, i\neq j}^l R^{\alpha~~\beta}_{~\mu_i~~\mu_j}\,f_{\mu_1\cdots \alpha\cdots \beta \cdots \mu_l} \ ,
\end{aligned}
\end{align}
where $R_{\alpha\beta\gamma\delta}$ is the Riemann curvature tensor on the coset space $G/H$.
This is a spinning wave equation for the tensor field $f$ of spin $l$ on the coset space $G/H$.
If we could find a Radon transform from the moduli space of conformal defects $\CM^{(d,m)}$ to the AdS space, the identity \eqref{Lichnerowicz} would be intertwined to the equation of motion for higher spin field on the AdS space.
Hence it is tempting to examine the defect OPE blocks for spinning defects and extend the bulk reconstruction program for higher spin fields.

\paragraph{Twist and anti-twist operators for entanglement entropy}
The calculation of entanglement entropy amounts to performing the path integral on a Euclidean manifold singular at a codimension-two entangling surface.
Such a computation can be reformulated as the path integral on a manifold without singularity, but with the insertion of a twist operator specifying an appropriate boundary condition on the entangling surface \cite{Calabrese:2004eu,Cardy:2013nua,Hung:2014npa,Bianchi:2015liz}.
For a spherical or planar entangling surface in CFT, the twist operator is a particular example of codimension-two conformal defects, whose dual defect becomes a pair of local operators located at the future and past tips $X_\pm$ of the causal diamond for the entangling surface (see also \cite{deBoer:2016pqk,Long:2016vkg}).

There are two types of codimension-two defects for a given spherical entangling surface, twist and anti-twist operators, depending on their orientations.
For example, if we associate twist operators to the entanglement entropy inside the sphere, anti-twist operators are to the entropy of the complementary region, namely the outside of the sphere.
This distinction does not matter as long as we are concerned with the entanglement entropy of a single spherical region, but a care must be taken when we deal with the entropy across two disjoint regions.
On the one hand the entanglement entropy can be represented as the correlation function of two twist operators, but on the other hand the entropy can be given, after an appropriate conformal transformation, by the correlator of twist and anti-twist operators associated to the complement of the spherical shell that is conformally equivalent to the two disjoint spherical regions.

In order to incorporate the difference between twist and anti-twist operators for entanglement entropy in our framework of conformal defects, we exploit the duality between a codimension-two defect and a pair of local operators and propose to introduce two types of local operators $\Phi(X_+)$ and $\tilde\Phi(X_-)$ describing the dual defect of a twist operator,
\begin{align}\label{DualMapEE}
	\CD^{(2)}_\text{twist}(P_\alpha) \quad \leftrightarrow \quad \CD^{(d)}_\text{twist}(
	\tilde P_{\tilde\alpha}) = \Phi(X_+)\, \tilde\Phi(X_-) \ .
\end{align}
With this refinement, the anti-twist operator can be distinguished from the twist operator by exchanging $\Phi \leftrightarrow \tilde\Phi$,
\begin{align}
	\CD^{(2)}_\text{anti-twist}(P_\alpha) \quad \leftrightarrow \quad \CD^{(d)}_\text{anti-twist}(
	\tilde P_{\tilde\alpha}) = \tilde\Phi(X_+)\, \Phi(X_-) \ .
\end{align}
In other words we assign to $\Phi$ and $\tilde\Phi$ different $\BZ_2$ charges with which we can determine the orientation of entangling regions.
This rule is consistent with the inversion map that flips the roles of twist and anti-twist operators.
It is intriguing to implement our proposal in the calculation of the entanglement entropy for a spherical shell as the correlation function of the four local operators by taking into account the $\BZ_2$ charges.

\paragraph{More general OPE blocks in defect CFT}
The method we used to derive the defect OPE blocks can be generalized similarly to the bulk-to-defect OPE and the OPE of defect local operators.
For the latter, we employ the spectral decomposition of the identity operator in the bulk CFT to deduce
\begin{align}
	\hat \CO_{i}(Y) &= \sum_n\,\int D^d X \,\langle \hat \CO_{i}(Y) \,\tilde\CO_n(X)\rangle\,\CO_n(X) \ ,
\end{align}
which reproduces the defect OPE when we choose the defect local operator to be the identity operator on the defect, $\hat \CO_i = \hat {\bf 1}$.
To achieve the bulk-to-defect OPE, we would rather use the spectral decomposition of the defect identity $\hat {\bf 1}$ in the defect theory, resulting in
\begin{align}
\CO_{i}(Y)|_\CD &= \sum_n\, \int_{\CD} D^d X \,\langle \CO_{i}(Y) \,\tilde{\hat{\CO_n}}(X)\rangle\, \hat\CO_n(X) \ ,
\end{align}
where the integration range is restricted on the support of the defect $\CD$.
These integral forms would shed light on new aspects of the OPEs in defect CFT and be worth further studies.

\paragraph{Codimension-one defect}
There is no dual object for a codimension-one defect, but one can still find a solution by extending the null cone $X^2=0$ to the whole embedding space.
The latter is regarded as the bulk AdS space and the dual defect is likely to sit on the tip of the causal diamond in the AdS space.
It is tempting to expect that the defect OPE block for $m=1$ measures the complexity of the vacuum state in CFT as both of them are calculated by integrating over the volume in the bulk \cite{Susskind:2014rva,MIyaji:2015mia}.\footnote{We thank K.\,Watanabe for discussion on this possibility.}


\acknowledgments  
We would like to thank M.\,Go, T.\,Ugajin, K.\,Watanabe and S.\,Yamaguchi for valuable discussion. 
The work of T.\,N. was supported in
part by JSPS Grant-in-Aid for Young Scientists (B) No.15K17628 and JSPS Grant-in-Aid
for Scientific Research (A) No.16H02182.
The work of M.\,F. was supported by JSPS fellowship for Young students.
The work of N.\,K. was supported in part by the Program for Leading Graduate Schools, MEXT, Japan and also supported by World Premier International Research Center Initiative (WPI Initiative), MEXT, Japan.

\appendix

\section{Defect OPE block of spinning primaries in the small radius limit}\label{ap:SmallRadius}
Generalizing the argument in section \ref{ss:IntegralRep} to the $l>0$ case is parallel to the $l=0$ case with a slight modification due to the tensor structure.
To this end, we introduce new vectors,
\begin{align}
	K_{\alpha\beta}(X) \equiv C_{P^\alpha P^\beta }\cdot X \ ,
\end{align}
whose norm for given indices $(\alpha,\beta)$ is
\begin{align}
	K_{\alpha\beta}(X)\cdot K_{\alpha\beta}(X) = (P^\alpha \cdot X)^2 + (P^\beta \cdot X)^2 \ .
\end{align}
The contraction of the indices gives
\begin{align}
	K^{\alpha\beta}(X)\cdot K_{\alpha\beta}(X) = 2 (P^\alpha \cdot X) (P_\alpha \cdot X) \ ,
\end{align}
with which we rewrite the correlator \eqref{One_Point_Defect} in the form,
\begin{align}\label{One_Point_Defect2}
	\langle \CD^{(m)}(P_\alpha)\, \CO_{\Delta, l}(X, Z)\rangle 
    = \frac{2^{(\Delta + l)/2}\,a_{\Delta, l}}{\left[ K^{\alpha\beta}(X)\cdot K_{\alpha\beta}(X)\right]^{(\Delta + l)/2}} \left[ (Z \cdot K^{\gamma\delta}(X))(Z \cdot K_{\gamma\delta}(X))\right]^{l/2}\ .
\end{align}
The vectors with $\alpha=m$ component dominate in taking the small radius limit:
\begin{align}
\begin{aligned}
	K_{\alpha m}(X) &~\longrightarrow ~\frac{1}{R}\, C_{P^\alpha C}\cdot X + O(1) \ ,&\quad & \text{for}\quad \alpha \neq m \ ,\\
    K_{\alpha \beta}(X) &~\longrightarrow ~O(1) \ ,&\quad & \text{for} \quad \alpha, \beta \neq m \ .
\end{aligned}
\end{align}
Substituting the correlator \eqref{One_Point_Defect2} into \eqref{Defect_OPE_Block} and taking the small radius limit yields the asymptotic behavior of the defect OPE block,
\begin{align}
\begin{aligned}
	\CB^{(m)}[P_\alpha, \CO_{\Delta,l}] &\sim R^\Delta \sum_{\alpha}\,\int D^d X \, \frac{1}{(C\cdot X)^{\Delta +l}}\,\tilde\CO_{d-\Delta, l} (X, C_{P^\alpha C}\cdot X) + \cdots \ ,\\
    &\sim R^\Delta\, \sum_{\alpha}\, \CO_{\Delta, l}(C,P_\alpha) + \cdots \ .
\end{aligned}
\end{align}
Note that the $\alpha=m$ term vanishes in the first line, but we take the summation for $\alpha$ over $\alpha = 1, \cdots, m$ so that the final form is invariant under the $SO(m)$ symmetry rotating the frame vectors.

\section{Proof of the inversion formula}\label{ap:Inversion}
\begin{figure}[t]
\centering
\begin{tikzpicture}
	\shade[left color=blue!80!green!30,right color=blue!15!green!5] 
  	(0,0) node[left] {\Large $\xi_p$} to [out=-10,in=150] (6,-2) -- (12,1) to [out=150,in=-10] (5.5,3.7) -- cycle;   
    
    \coordinate (a) at (7, 5);
    \coordinate (b) at (7, 1.5);
    \coordinate (c) at (4, 0);
    
    \draw[fill=black] (a) circle (2pt) node [above] {\Large $x$};
    \draw[fill=black] (b) circle (2pt) node [right] {\Large $x_0$};
    \draw[fill=black] (c) circle (2pt) node [left] {\Large $y$};
    
    \draw[] (a) -- node[midway, above right] {\Large $p$} (b);
    \draw[] (a) -- node[midway, left] {\Large $q$} (c);
    \draw[] (c) to [out=30, in=195] node[midway, below] {\Large $r$} (b);
\end{tikzpicture}
\caption{Trigonometry for a triangle in a hyperbolic space}
\label{fig:Hyp_trig}
\end{figure}
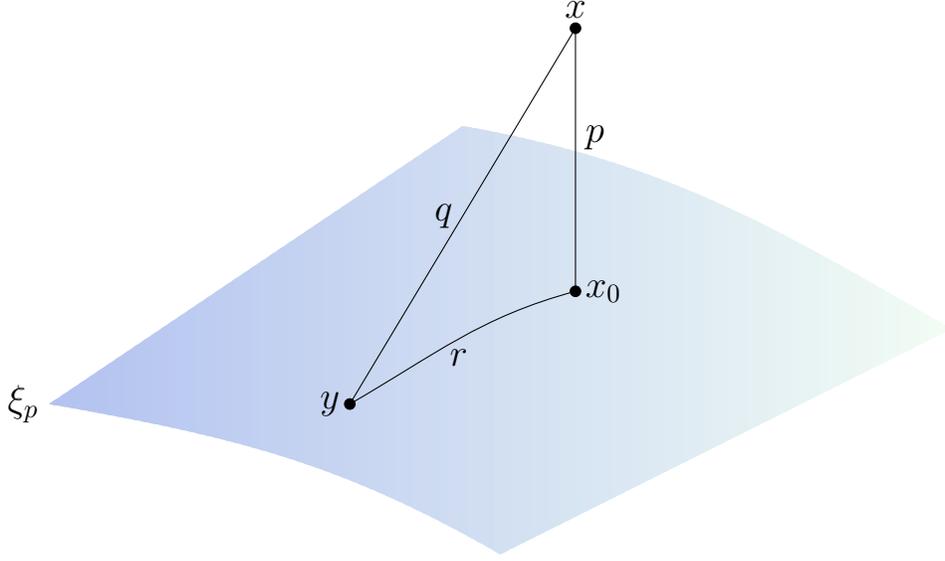

The inversion formula \eqref{inversion_formula}, which played the central role in section \ref{ss:HKLL}, can be proven as follows \cite{helgason}.
We fix a point $x$ in $\mathbb{H}^{d+1}$ on which we want to reconstruct the value of a function $\phi$. 
Let $x_0$ be the nearest point in a $k$-dimensional geodesic submanifold $\xi_p$ from $x$, $r$ be the distance between $x_0$ and a point $y$ on the $\xi_p$, and $q$ be the distance between $x$ and $y$ as shown in figure \ref{fig:Hyp_trig}.
The distances between these points satisfy the triangle relation,
\begin{equation}
\cosh q = \cosh p \ \cosh r\,.
\end{equation}

We rewrite the mean-value operator $(M^{p} \hat{\phi}) (x)$ as, 
\begin{align}
\begin{split}
(M^{p} \hat{\phi}) (x) &= \int_{K} \hat{\phi}(gk\cdot\xi_p)\,dk =  \int_{K} \int_{y\in\xi_p} \phi (gk\cdot y )\,dm(y)\, dk\ ,\\
&= \int_{y\in\xi_p} (M^{q} \phi)(x) \,dm(y)\ ,
\end{split}
\end{align}
with $q = d(x,y)$ and $(M^{q} \phi)(x)$ appeared in the last equality that is defined as a mean-value operator on a surface at distance $q$ from a point $x$,
\begin{align}
	(M^{q= d(x,y)} \phi)(x = g\cdot o) \equiv \int_K\, \phi (gk\cdot y)\, dk \ .
\end{align}
Recall that $(M^{p} \hat{\phi}) (x)$  only depends on the radial direction, and it enables us to see $M^{p}\hat{\phi}$ as a function of $r$.
In the global coordinate of $\mathbb{H}^{d+1}$, the integral over the points in $\xi_p$ can be represented in the polar coordinate, and thus we have
\begin{equation}\label{inversion_integrand}
(M^{p} \hat\phi)(x) = \Omega_k \int_0^{\infty} (M^{q} \phi)(x)\, \sinh ^{k-1}r\,dr\,,
\end{equation}
with $\Omega_k$ is the area of the surface at distance $r$ from $x_0$, restricted to the $k$-dimensional geodesic submanifold.
In order to indicate that $ (M^{q} \phi)(x)$ is the only function of the distance $q$ and $(M^{p} \hat{\phi}) (x)$ is a function of $p$, we denote them by
\begin{equation}
 F(\cosh q) = (M^{q} \phi)(x)\ , \qquad \hat{F}(\cosh p) = (M^{p} \hat\phi)(x)\ .
\end{equation}
By changing variables as $\cosh p \to t, \cosh r \to s$,  \eqref{inversion_integrand} is written as
\begin{equation}
\hat{F}(t) = \Omega_k\,\int_1^{\infty}F(ts)(s^2-1)^{(k-2)/2} \ ds\ .
\end{equation}
Inverting this equation leads to
\begin{equation}
r^{-1}F(r) = -c_k \left(\frac{d}{dr^2}\right)^{k}\int_r^\infty t^{k}(t^2-r^2)^{(k-2)/2}\hat{F}(t)\,dt\,.
\end{equation}
The left hand size reduces to $\phi(x)$ when $r=1$ as it becomes the average over a surface of zero radius: $F(1) = (M^{d(x,y)=0}\phi)(x)$.
Thus the inversion formula \eqref{inversion_formula} is obtained by letting $r=1$.

\section{Construction of a bulk scalar operator in $d=2$}\label{ap:HKLL}
In this section, we reproduce the result of \cite{Czech:2016xec}, that is the HKLL formula in the AdS$_3$ space.
This corresponds to the $d=2$, $m = 2$ case.
For this purpose, we begin with introducing the global section in the embedding space formalism.

A vector in the global section, constrained on the null cone, is in general, given by \cite{kaplan2015lectures}, 
\begin{equation}\label{global_section}
	X = (e^t,\, e^{-t},\, \cos \theta,\, \sin \theta)\ .
\end{equation}
Let us briefly explain why this is so.
We choose the global coordinates of AdS$_{d+1}$ (with the defining relation $-Y^+ Y^- + \sum_i (Y^i)^2 = -1$) as
\begin{equation}
	(Y^+, Y^-, Y^i)  = (e^\tau/\cos \rho,\, e^{-\tau}/\cos \rho,\, \tan\rho\, \Omega_i) \ ,
\end{equation}
where $\Omega_i$'s are the vectors on the $d$-dimensional sphere.
In these coordinates, the boundary corresponds to the $\rho \to \pi/2$ limit.
That is, we use the infinitesimal parameter $\epsilon$, and substitute  approximation $\rho = \pi/2 -\epsilon f(\tau, \Omega)$ for some function $f(\tau, \Omega)$.
In order to reproduce (\ref{global_section}), we choose $f(\tau, \Omega) = e^{-\tau}$.
By rescaling $X \equiv \epsilon\, Y$, we obtain $X = (e^{2\tau}, 1, e^\tau\Omega_i)$. 
Dividing by $e^{\tau}$ gives us the result \eqref{global_section} as the null vector dose not depend on the over all rescaling.

\begin{figure}[t]
\centering
\begin{tikzpicture}[scale=.7]
\tkzInit[xmin=-4.1,xmax=5.2,ymin=-4.1,ymax=6]
\tkzClip[space=.5]
\tkzDefPoint(70:6){A}
\tkzDefPoint(0,0){O} \tkzDefPoint(4,0){R}
\tkzDrawCircle(O,R)
\tkzDefPoint(70:4){D}
\tkzTangent[from = A](O,R) \tkzGetPoints{B}{C}
\tkzDrawArc[color=blue,thick](A,B)(C)
\tkzDrawSegment[style=dashed](O,D)
\tkzDrawSegment[style=dashed](O,R)
\tikzset{compass style/.append style={<->}}
\tkzDrawArc[color=orange,style=double](O,C)(D)
\tkzLabelAngle[0.5,right](O,D,C){\large $\alpha$}
\tkzMarkAngle[fill=blue!25,mkpos=.2, size=0.7](R,O,D)
\tkzLabelAngle(R,O,D){\large $\theta_{c}$}
\tkzDrawPoints(O)
\tkzLabelPoints(O)
\node[above right] at (C) {\large $X_1$};
\node[above left] at (B) {\large $X_2$};
\node at (-1.15,2) {\large $\xi$};
\end{tikzpicture}
\caption{A pair of points at $X_1$ and $X_2$ as a codimension-two defect on a circle at a time slice, and the horocycle $\xi$ in $\BH^2$ ending on them at the boundary}
\label{fig:HKLL_d=3}
\end{figure}
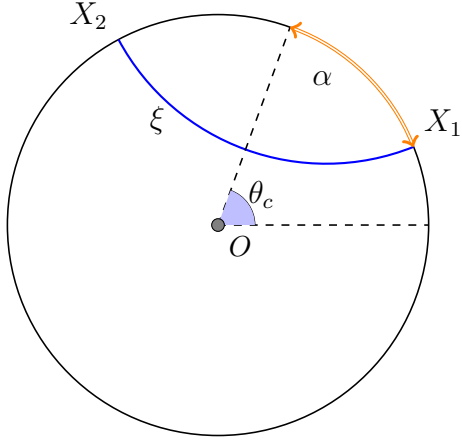

In order to construct an AdS scalar field at the origin of $\BH^2$, we pick up a horocycle $\xi$ and read off the position of the corresponding defect following the steps 1 and 2 in section \ref{ss:HKLL}.
We assume horocycle and codimension-two defect are constrained on the time slice, $t=0$.
As clear from figure \ref{fig:HKLL_d=3}, it is enough to specify two points $X_{1,2}$ on the circle parameterized by $\theta$ for fixing a horocycle in the time slice $\BH^2$.
Thus we choose the two points as follows:
\begin{equation}
	X_1 = \left(1, 1,\ \cos(\theta_c-\alpha),\ \sin(\theta_c-\alpha)\right)\ ,\qquad X_2 = \left(1, 1,\ \cos(\theta_c+\alpha),\ \sin(\theta_c+\alpha)\right)\ .
\end{equation} 
Here $\theta_c$ and $\alpha$ are the center and the radius of the defect respectively.

Since the defects are constrained on the time slice, one of the frame vector $P_1$ has to be parallel to the time direction at $t=0$,
\begin{equation}
	P_1 = (1, -1, 0, 0)\ .
\end{equation}
Note that this reduces the defect moduli space to the subspace $\CM^{(1,1)}$.
It follows that the other frame vector $P_2$ is determined to be 
\begin{equation}
	P_2 = (\cot \alpha,\ \cot \alpha,\ \cot \theta_c \csc \alpha,\ \sin\theta_c\csc \alpha)\ .
\end{equation}
This achieves the step 3.
Now we turn to the step 4.
To compare with the result in \cite{Czech:2016xec}, we use the defect OPE block of the form \eqref{Defect_OPE_Block_flipped}, which reduces in the present case to
\begin{align}
\begin{aligned}\label{dm2block}
	\CB^{(2)}&[P_\alpha, \CO_{\Delta}] \\
    	&= \frac{a_{2-\Delta}}{\CN_{2-\Delta}} \int_{-\infty}^\infty dt \int_0^{2\pi}d\theta\ \ \left[2\frac{\left(\cosh t-\cos(\theta-\theta_c - \alpha)\right)\left(\cosh t-\cos(\theta-\theta_c + \alpha)\right)}{1-\cos(2\alpha)}\right]^{\frac{\Delta-2}{2}}\CO_{\Delta}(t, \theta)\ .
\end{aligned}
\end{align}
It is notable that the inversion formula (\ref{Bulk_reconstruction}) simplifies for $d=m=2$,
\begin{align}
\begin{split}
\phi (x) & = -c_{1} \left[ \frac{d}{d (r^2)}\, \int_r^\infty dt\, (t^2 - r^2)^{-1/2 }t\, (M^{s(t)} \hat\phi) (x) \right]_{r=1} \ ,\\
&= c_{1} \left[ \frac{d}{d (r^2)}\, \int_r^\infty dt\, (t^2 - r^2)^{1/2 } \frac{d}{dt}(M^{s(t)} \hat\phi) (x) \right]_{r=1}\ ,\\
&= -c_{1}/2  \int_1^\infty dt\, (t^2 - 1)^{-1/2 } \frac{d}{dt} (M^{s(t)} \hat\phi) (x) \ ,\\
&=-\frac{1}{\pi} \int_0^{\infty} dp\  \frac{1}{\sinh p}\frac{d}{dp} (M^{p} \hat\phi) (x)\ .
\end{split}
\end{align}
From the first to second line, the integration by parts was carried out for $(t^2 - r^2)^{-1/2 }t = \partial_t (t^2 - r^2)^{1/2 }$, and from the third to forth line, we introduced the new variable $p$ by $t = \cosh p$.
Putting all things together, we arrive at the final form,
\begin{align}
\begin{split}
	\phi(0) = -\frac{a_{2-\Delta}}{\pi\,\CN_{2-\Delta}} \int_{-\infty}^\infty dt \int_0^{2\pi}d\theta\ \ K_\Delta(t)\,\CO_{\Delta}(t, \theta)\ ,
\end{split}
\end{align}
with the kernel given by
\begin{equation}
\ K_\Delta(t) =  \int_{0}^{\pi/2} d\alpha \tan \alpha \frac{d}{d\alpha} \ \int_{-\pi/2}^{\pi/2} d\theta_c\ \left[2\frac{\left(\cosh t-\cos(\theta-\theta_c - \alpha)\right)\left(\cosh t-\cos(\theta-\theta_c + \alpha)\right)}{1-\cos(2\alpha)}\right]^{\frac{\Delta-2}{2}}\,.
\end{equation}
Note that the integration over $\theta_c$ comes from the group integration on $SO(2)$, and we used the relation $p = \text{arccosh} (1/\sin \alpha)$, derived from \eqref{rel_t_R}.
This agrees with the kernel given by \cite{Czech:2016xec}, up to the Wick rotation $t \to -i t$.
The double integrations in the kernel were carried out there, which in our case results in the Euclidean version of the HKLL formula \cite{Hamilton:2006fh} as expected,
\begin{equation}
	K_\Delta(t) = \frac{2^{\Delta-2}(\Delta-1)}{\pi^2}(\cosh t)^{\Delta-2} \log\cosh t\ .
\end{equation}

\bibliographystyle{JHEP}
\bibliography{Defect_OPE_Block}

\end{document}